\documentclass[3p]{elsarticle}

\usepackage{lineno,hyperref}
\usepackage{amsmath,amssymb,amsfonts}
\usepackage{algorithmic}
\usepackage{algorithm}
\usepackage{multirow}
\usepackage{multirow}
\usepackage{graphicx}
\usepackage{textcomp}
\usepackage{cellspace}
\usepackage{wrapfig}
\usepackage{url}
\usepackage{color}
\newtheorem{myDef}{Definition}
\newtheorem{myTheo}{Theorem}

\usepackage{tabularx}
\usepackage[dvips]{epsfig}
\usepackage{lscape}
\usepackage[caption=false,font=footnotesize]{subfig}
\modulolinenumbers[5]
\newcommand{\tabincell}[2]{\begin{tabular}{@{}#1@{}}#2\end{tabular}}

\journal{Big Data Mining and Analytics}

\begin{document}

\begin{frontmatter}

\title{Towards Blockchain-Assisted Privacy-Aware Data Sharing For Edge Intelligence: A Smart Healthcare Perspective}



\author[mymainaddress]{Youyang Qu}
\author[secondaddress]{Lichuan Ma}
\author[mythirdaddress]{Wenjie Ye}
\author[myfourthaddress]{Xuemeng Zhai}
\author[myfifthaddress]{Shui Yu}
\author[mysixthaddress]{Yunfeng Li \corref{mycorrespondingauthor}}
\author[myfirstaddress]{David Smith}

\address[mymainaddress]{Data61, CSIRO, Australia}
\address[mysecondaryaddress]{Xidian University, China}
\address[mythirdaddress]{Victoria University, Australia}
\address[myfourthaddress]{University of Electronic Science and Technology of China, China}
\address[myfifthaddress]{University of Technology Sydney, Australia}
\address[mysixthaddress]{CNPIEC KEXIN LTD, China}





\begin{abstract}

The popularization of intelligent healthcare devices and big data analytics significantly boosts the development of smart healthcare networks (SHNs). To enhance the precision of diagnosis, different participants in SHNs share health data that contains sensitive information. Therefore, the data exchange process raises privacy concerns, especially when the integration of health data from multiple sources (linkage attack) results in further leakage. Linkage attack is a type of dominant attack in the privacy domain, which can leverage various data sources for private data mining. Furthermore, adversaries launch poisoning attacks to falsify the health data, which leads to misdiagnosing or even physical damage. To protect private health data, we propose a personalized differential privacy model based on the trust levels among users. The trust is evaluated by a defined community density, while the corresponding privacy protection level is mapped to controllable randomized noise constrained by differential privacy. To avoid linkage attacks in personalized differential privacy, we designed a noise correlation decoupling mechanism using a Markov stochastic process. In addition, we build the community model on a blockchain, which can mitigate the risk of poisoning attacks during differentially private data transmission over SHNs. To testify the effectiveness and superiority of the proposed approach, we conduct extensive experiments on benchmark datasets.

\end{abstract}


\end{frontmatter}


\section{Introduction}

With recent advances like machine learning and intelligent edge devices, the wide proliferation of smart healthcare systems has been enabled. Consequently, a wide range of applications has emerged and serviced our daily life. Among all of them, smart health networks (SHNs) is one of the most widespread services that has been adaopted in real-world scenarios \cite{garcia2017leaking}. Healthcare has been a long-lasting concern of the society, and the development of advanced technologies takes it to a new stage. In this case, people rely more and more on smart healthcare services to enhance their living quality. To make this happen, doctors or patients are willing to establish communities in SHNs with regards to a certain decease, for example, Doximity and Curofy \cite{DBLP:journals/cm/HossainXLBE18}\cite{wang2021security}. It is worth mentioning that SHN users are likely to form the almost same communities in various smart healthcare networks \cite{DBLP:journals/tkde/ZhouLZM16} \cite{DBLP:journals/iotj/CatarinucciDMPP15}. On one hand, more services may be delieved. But on the other hand, multiple data resources of uses are natually disclosed.

Intuitively, the key target of SHNs is to share useful information among community uses. The shared health data usually contains texts, medias, as well as spatial and temporal data \cite{DBLP:journals/access/Yu16}\cite{qu2018privacy}. The combination of these data can be used to re-identify a specific person and leads to further privacy disclosure. Thus, great risks are raised when sensitive health data of individuals are published without proper pre-processing. This can be even worse when different parties can access the data without proper access control \cite{DBLP:journals/cm/HeYCGX18}\cite{nie2022blockchain}.

It has been agreed on that sensitive information, especially sensitive health information raises the financial interest of diverse adversaries or attackers. New attacks are reported every several months, or even weeks. Adversaries are usually patient and smart enough to collect individual's data from various data sources, which can be used for re-identification \cite{DBLP:conf/kdd/ZhangTYPY15}. This also causes a worse situation that linkage attacks targeting on further private data is possible \cite{DBLP:series/asc/ZhangZY16} \cite{DBLP:journals/tdp/Merener12}. Therefore, it is necessary to establish effective privacy protection mechanisms for SHNs.

To preserve privacy, three classic methods have been well studied, which are cryptography, anonymization and clustering, and differential privacy. Cryptography preserves privacy during the data packet transmission process but can hardly preserve privacy against data recipients \cite{YCYJ2016}. Anonymization and clustering have been developing for several decades. Several benchmark methods include $K$-anonymity \cite{PL1998}, $L$-diversity \cite{ADJM2007}, and $T$-closeness \cite{NTS2010}. Existing clustering methods consider record number, record type, record distribution, or a combination of them. However, there are not suitable for streaming data sharing. Differential privacy \cite{DBLP:conf/icalp/Dwork06}\cite{DBLP:reference/crypt/Dwork11} is a powerful privacy protection tool constrained by mathematical theories. But for classic differential privacy and its variants, the privacy protection level is usually constant. 

There are some pioneering works in personalized privacy protection. For instance, using virtual online distance as the penalization index is a representative work \cite{DBLP:diffusion_personalized}. However, virtual online distance has some issues during deployment. First, the distance is not easy to define. Second, friends in the network may have the exactly same distance as attackers in some cases. Besides, some blockchain-based solutions are devised to potentially add extra protection for privacy. For example, Wang et al. developed a blockchain-powered healthcare system \cite{DBLP:journals/tcss/WangWWQYOGW18}. Blockchain-based solutions can ensure authentication and integrity, but the public accessibility of health data put privacy at great risk.

Motivated by this, we developed a personalized differential privacy protection model, which can derive an optimized trade-off between privacy protection and health data utility. Personalization is achieved by a trust level measured by community density. By defining community density, it can be used as the measurement of the intimacy of a group of people. To avoid utility loss, we devise a novel community partition method based on the work done by Ahn et at \cite{ahn2010link}. Besides, we use a semi-sigmoid function as a mapping function, which maps the community-enabled trust to a protection level. Then, a Markov stochastic process is built to uncouple the randomized noise relationship, mitigating the linkage attacks. Moreover, we design a tailor-made blockchain structure to accommodate the community-based personalized privacy protection model while avoiding any data falsification attacks and ensuring data integrity during transmission over SHNs.

The main contributions of this work are summarized as follows.

\begin{itemize}

\item[$\circ$] \textbf{Personalized and Trustworthy Privacy Protection}: We define community density to measure the trust within communities. Then, the trust is mapped to a privacy protection level constrained by differential privacy. In this way, we develop a novel personalized and trustworthy privacy protection model.

\item[$\circ$] \textbf{Data Falsification Proof}: We devise a tailor-made blockchain structure that can support the personalized and trustworthy privacy protection model. The differentially private health data is guaranteed to be authenticated provided by the features of this blockchain structure.

\item[$\circ$] \textbf{Attack-Proofing and Optimization Trade-off}: We properly decouple the data correlation with Markov stochastic process. Therefore, the linkage attack can be eliminated. Furthermore, the proposed model achieves an optimized trade-off between personalized privacy protection and improved data utility.

\item[$\circ$]  Extensive results obtained from experiments show that the proposed system can achieve a good balance between personalized privacy protection and health data utility. Besides, the system can defeat leading attacks like linkage attacks and poisoning attacks.

\end{itemize}

The remainder of this article is outlined as follows. In Section \ref{sec:related_works}, we discuss existing works of privacy preservation in SHNs. Then, the personalized and trustworthy privacy protection model as well as the tailor-made blockchain system are described in Section \ref{sec:system_modelling} and Section 4. After that, in Section \ref{sec::attack_proof}, we present the noise decoupling mechanism designed for optimized privacy-utility balance, followed by performance evaluation in Section \ref{sec:experiment}. In the end, we provide conclusion remarks in Section \ref{sec:summary}.

\section{Related Works}
\label{sec:related_works}

Research related to privacy preservation in Smart Healthcare Networks (SHNs) has gained significant attention due to the sensitive and personal nature of healthcare data. Privacy preservation techniques aim to protect individuals' privacy while enabling the sharing and analysis of healthcare data. These approaches offer several advantages, but also face certain shortcomings.

One major advantage of privacy preservation techniques in SHNs is the protection of individuals' sensitive medical information. By applying privacy-preserving mechanisms, such as data anonymization, encryption, or differential privacy, personally identifiable information can be safeguarded. This ensures that unauthorized entities cannot directly link healthcare data to specific individuals, reducing the risk of privacy breaches.

Another advantage is the potential to enable secure data sharing and collaboration among healthcare stakeholders. Privacy-preserving techniques allow healthcare providers, researchers, and organizations to share data while maintaining privacy. This promotes collaborative efforts in research, clinical decision-making, and public health without compromising sensitive information.

Furthermore, privacy preservation techniques contribute to building trust and compliance with privacy regulations. Patients and individuals are more likely to participate in data sharing initiatives if they have confidence in the privacy protections implemented. Compliance with regulations, such as the Health Insurance Portability and Accountability Act (HIPAA) in the United States or the General Data Protection Regulation (GDPR) in the European Union, is crucial in maintaining ethical and legal standards.

However, privacy preservation in SHNs also has some shortcomings. One significant challenge is balancing privacy protection with data utility. Applying rigorous privacy measures may introduce noise or limitations to the data, which can potentially impact its usefulness for analysis and decision-making. Striking the right balance between privacy and utility remains an ongoing research challenge.

Another issue is the potential for re-identification attacks. Despite privacy-preserving techniques, there is still a risk of re-identification when data is combined or linked with other external information sources. Techniques and safeguards need to be continuously developed and updated to address this vulnerability.

Additionally, the complexity and diversity of healthcare data pose challenges for privacy preservation. Healthcare data encompasses a wide range of data types, including structured medical records, genomics data, and wearable sensor data. Developing privacy mechanisms that are effective across diverse data types and modalities requires ongoing research and adaptation.

Overall, research in privacy preservation for SHNs is crucial for protecting sensitive healthcare information, facilitating secure data sharing, and ensuring compliance with privacy regulations. While these techniques offer notable advantages, addressing the challenges of maintaining data utility, preventing re-identification attacks, and accommodating diverse healthcare data types will be vital for advancing privacy preservation in SHNs.

Four successful branches of privacy protection solutions include clustering, cryptography, game theory, and differential privacy \cite{qu2022fedtwin}. The clustering-based solutions focus on how to group the datasets considering the consistency of magnitude, diversity, and distribution between each cluster and the whole datasets \cite{PL1998}\cite{ADJM2007}\cite{NTS2010}. However, such a clustering approach fails to function well for large-scale and heterogeneous datasets. Cryptography-based solutions protect privacy in a point-to-point manner. However, an unknown adversary always has the potential to decrypt the data while the computational complexity is quite high \cite{YCYJ2016}. Game theory is another potential optimization approach so as to balance privacy with data utility. However, such an approach also suffers from the modelling accuracy of actions and payoffs in different contexts. It is worth noting here that the increasing complexity of the number of participants makes it hard to generalize \cite{DBLP:journals/ton/WangZ16}. For classic differential privacy and its variants, there are mathematical constraints to explain the privacy protection performance, but the privacy protection level is usually constant \cite{DBLP:conf/icalp/Dwork06}\cite{DBLP:conf/eurocrypt/DworkKMMN06}\cite{DBLP:reference/crypt/Dwork11}\cite{pokhrel2020qos}.

In the privacy protection domain, linkage attacks have always been a big problem, especially when we consider multiple SHNs. A lot of features have been used for accurate re-identification in this scenario, such as content matching, profile matching, unique structure matching, etc \cite{DBLP:journals/access/MaQHHWSZW17}\cite{pokhrel2020privacy}. 

By using binary classifiers, Perito et al. compared the similarity of pseudo identities and achieved re-identification of users in \cite{DBLP:conf/pet/PeritoCKM11}. While in \cite{DBLP:conf/kdd/ZafaraniL13}, Liu et al. defined behaviors based on pseudo identities to establish a user mapping model. With media data, Xu et al. devised a high-efficiency identification model as well as a countermeasure in \cite{DBLP:journals/tdsc/XuGGFL17}. In addition, both static location data and trajectory data are utilized to re-identify a specific individual in \cite{DBLP:journals/tsc/SchlegelCHW17} and \cite{miller2018exploratory}, respectively. Based on existing research, Li et al. surveyed the issues and solutions of de-anonymization and aggregation of heterogeneous social networks.

To preserve the data privacy of SHNs, existing research has made great efforts, especially from the aspect of health data. Yang et al. leveraged access control to preserve the privacy of health data sharing over the Internet of Things \cite{DBLP:journals/isci/YangZGLC19}. Another similar research is performed by Zhang et al., who further introduced attribute-based access control  to preserve health data privacy while considering the efficiency \cite{DBLP:journals/iotj/ZhangZD18}. In the smart wearable healthcare devices case, Liu et al. designed a cooperative privacy-preserving model \cite{DBLP:journals/iotj/LiuYYN19}. The above three are cryptography methods. Despite their effectiveness, the low efficiency and granularity of privacy protection prevent their further application in SHNs.

In SHNs, an emerging technology, blockchain, is also considered by researchers \cite{qu2022blockchain}. To preserve big health data privacy, Xu et al. designed a blockchain-based privacy protection model \cite{DBLP:journals/iotj/XuXLTHHY19}. Decentralized privacy protection of SHNs data is considered by Dwivedi et al. using a tailor-made healthcare blockchain \cite{dwivedi2019decentralized}. Besides, Peterson et al. also proposed blockchain-based private SHN data-sharing schemes in \cite{peterson2016blockchain}. However, for most blockchain-based systems, the feature of publicly accessible private data is still a big issue and not well-addressed \cite{zubaydi2019review}.

In healthcare application scenarios, differential privacy has been widely deployed. Nevertheless, the constant privacy protection level limits its further development. Therefore, a more flexible protection mechanism like personalized differential privacy is necessary. But personalized privacy suffers from security and privacy vulnerabilities as well. This is even worse in the smart healthcare context since the data are highly confidential and related to physical security. Based on our literature review, this part has not yet been well-discussed. The idea we propose in this paper is a preliminary exploration in this field that integrates personalized differential privacy protection, trust, blockchain, etc., to provide several advantageous features.

\section{Personalized and Trustworthy Privacy Protection in SHNs}
\label{sec:system_modelling}

In this section, we present the devised personalized and trustworthy privacy-preserving model for data sharing in SHNs. We first describe our community detection algorithm and discuss the structure-based parameters. In this paper, we used a modified link community algorithm to achieve personalized privacy protection. The major modification is that we require users to belong to a single community at the final stage. The existing link community algorithm allows the user to be a part of several communities, which is not feasible in this scenario. That is because a user may access a piece of data but under different protection levels if he/she is part of several communities. As will be analyzed further, linkage attacks can be launched by a single user without collusion, which often provides incentives to malicious users. 
To overcome this problem, each user is only allocated to one single community with the highest trust level (highest community density and lowest protection level).

Then, to map the trust level to the privacy protection level, we introduce a semi-sigmoid function. The function value barely increases in the low range and high range but increases almost linearly in the middle range. After that, personalized and trustworthy privacy protection based on trust level is discussed and explained in detail.

\subsection{The Modeling of SHN's Graph Structure}

In recent years, SHNs have been evolving at a faster pace. They are of several different forms. For example, some of them are tailor-made SHNs for doctors or patients only to share the disease information like diabetes or cancers, such as Doximity and Curofy \cite{DBLP:journals/cm/HossainXLBE18}. Besides, some other social networks provide smart healthcare services, which is a sub-network functioning within the whole social network, such as Facebook. No matter the form of the SHNs, they are essentially networks with nodes as the users and edges denoting users' relationships. Therefore, it is reasonable to use the undirected graph of graph theory to model an SHN.

In this context, we use a single SHN as an example. Multiple SHN privacy issues is a simple extension of the single SHN. As mentioned above, the undirected graph is used to model the SHN, which is the foundation of the whole system. A single SHN is denoted by $G$ where $G = \{u_i, e_i, c_i \ | \  u \in U, e \in E, c \in C\}$. In this graph, $u \in U$ is the node (user in SHN), $e\in E$ is the edge (relationship between two users of SHN), and $c\in C$ is the community (formed by multiple users in SHN).

If there exists one edge (e.g. $\{e_{i,i+1}, ..., e_{i+1,i+n}, e_{i+n,j}\}$) between two nodes $(u_i, u_j)$, then there is relationship between these two nodes. In this context, different from traditional methods, the community is defined over a set of edges instead of nodes, which is $C = \{e_{i,j} |e \in E\}$.

In order to better clarify, we assume the modeled $G$ to be an undirected graph. The proposed model functions well even if this assumption is removed. In addition, we assume there is no trusted central authority. The decentralized blockchain system processes the data with $\epsilon$-differential privacy and delivers data with privacy-preserving communication. The total privacy budget of an SHN is set to be $B$. To allocate the budget to each user, it depends on the personzalized sensitivity value modeled in the following sections.

Privacy losses accumulate with an increasing number of data sharing. When two answers are responded to an individual, the total privacy loss increases while the privacy protection weakens. To guarantee significant privacy protection, the data curator should set a maximum privacy loss, in particular, $\epsilon$, which is also known as the privacy budget. For instance, a privacy ``cost'' incurs when data is shared under a defined privacy protection level. The continuous data-sharing process results in accumulating privacy loss. In Eq. \ref{e_differential_privacy}, it can be told that $\epsilon$ is the power of the natural logarithm. Thus, the protection level will be promoted with the decrease of $\epsilon$. Data utility will decrease in the meantime.

In Figure~\ref{f_mobile_social_network}

\subsection{Community Structure Detection}

To personalize the privacy protection level, we use community density to evaluate the trust within the community. The community density method is more suitable in this scenario compared with the traditional method like virtual online distance. The virtual distance only considers the relationship between two users, and may not be practical when attackers are within a relatively short distance. However, for communities, we evaluate the interaction among all members, which makes the trust more reliable.

The trust among users is evaluated by community density. If the users have dense interaction with each other, the trust in the community is high and thereby privacy protection will be released correspondingly. This is because users may share more information with people they trust, even in online scenarios like SHNs. Based on the density value, we map it to customizable privacy protection levels. To calculate community density, we first partition the whole graph into communities using Algorithm~1.

\renewcommand{\algorithmicrequire}{\textbf{Input:}} 
\renewcommand{\algorithmicensure}{\textbf{Output:}}
\begin{algorithm}[!h]  
    \label{algorithm1}
    \caption{Edge Pairs based Community Detection}
    \begin{algorithmic}[1]
    \REQUIRE A smart healthcare network graph represented by $G$;
    \ENSURE A set of detected communities $C = \{P_1, P_2, ..., P_c\}$;
    	\STATE Initialize each edge $e_i \in E$ as a community;
	\STATE Initialize amount of neighbours to each end node $n_+(i)$;
	\STATE Calculate edge pair similarity $S\Big(e_{ij},e_{ik}\Big)$ with $n_+(j)$ and $n_+(k)$ ;
	\STATE Sort all similarities $S\Big(e_{ij},e_{ik}\Big)$ in descending order;
	\STATE Merge the communities based on the ordered edge pairs;
	\STATE Represent the merging process with a tree structure;
	\STATE Define a threshold for partitioning density $D_t$
        \WHILE {$D < D_t$}
        \STATE Calculate edge pairs number $M$ and communities number $|C|$;
        \STATE Calculate $m_c$ and $n_c$ as edge pair number and node number in a community $P_c$;
        \STATE Derive partitioning density of $P_c$ with $m_c$ and $n_c$;
        \STATE Update partitioning density of whole smart healthcare network as $D = \frac{1}{M} \sum_c m_c D_c$;
        \ENDWHILE 
        \STATE Transform edge pair-based tree graph into node-based graph;
        \STATE Identify communities $C = \{P_1, P_2, ..., P_c\}$ and overlapped nodes $u_i \in U$;
        \STATE Output the communities $C = \{P_1, P_2, ..., P_c\}$.
    \end{algorithmic}  
\end{algorithm}

Algorithm~1 presents an algorithm for detecting communities in a smart healthcare network. Initially, each edge in the network is treated as an individual community. By examining the connectivity patterns, the algorithm calculates the similarity between pairs of edges, considering the number of neighbors of the connected nodes. These similarities are then sorted in descending order. Starting from the pair with the highest similarity, the algorithm progressively merges the corresponding communities, representing the merging process using a tree structure. The iteration continues until the partitioning density, which measures the density of connections within communities, reaches or exceeds a predefined threshold. The algorithm transforms the edge pair-based tree graph into a node-based graph, where each node represents a community. Finally, it identifies the communities and any overlapped nodes within the transformed graph, which are then outputted as the detected communities in the smart healthcare network.

The algorithm begins by initializing each edge in the smart healthcare network graph G as a separate community. This initialization step has a time complexity of O($|E|$), where $|E|$ represents the number of edges in the graph. Similarly, the initialization of the number of neighbors to each end node also takes O($|E|$) time as it requires iterating over all the edges in the network.

The next steps involve calculating the similarity between pairs of edges and sorting them in descending order based on these similarities. The time complexity of calculating the edge pair similarity depends on the specific method used and can range from O(1) to O($|E|^2$). Sorting all the similarities takes O($|E|^2 log |E|$) time in the worst case, assuming a comparison-based sorting algorithm like quicksort or mergesort.

The algorithm then proceeds to merge the communities based on the ordered edge pairs. The time complexity of this step depends on the specific merging method used and can range from O(1) to O($|E|$) depending on the merging strategy and any additional computations involved. Representing the merging process with a tree structure also takes O($|E|$) time.

Next, the algorithm enters a while loop that continues until a partitioning density threshold is reached. The number of iterations in the loop can vary depending on the network's characteristics and convergence behavior. In the worst case, the loop can have a time complexity of O($|E|$) if each iteration involves computations that scale linearly with the number of edges.

The transformation of the edge pair-based tree graph into a node-based graph requires traversing the tree structure and constructing the new graph representation. The time complexity of this step can range from O($|E|$) to O($|V|$), where $|V|$ represents the number of nodes in the resulting node-based graph. Identifying communities and overlapped nodes depends on the specific method used and can range from O($|V|$) to O($|E|$) in time complexity.

In summary, the overall time complexity of the algorithm ranges from O($|E|^2 log |E|$) to O($|E|$) in the worst case, depending on the specific methods used for similarity calculation, merging, and community identification. The input size and the specific characteristics of the smart healthcare network graph heavily influence the algorithm's computational complexity.

Different from traditional node-based communities, we use a set of edges to represent the community as $C = \{e_{i,j} |e \in E\}$. In the set of edges, each edge should be linked to at least one other edge in this community. That means no independent edge is allowed. In this way, we can avoid a node in multiple communities at the same time. This is because of the edge-based community detection algorithm. The ``overlapped nodes problem'' is thereby addressed.

\begin{figure}[!htbp]
\centering
\includegraphics[width=2.4in]{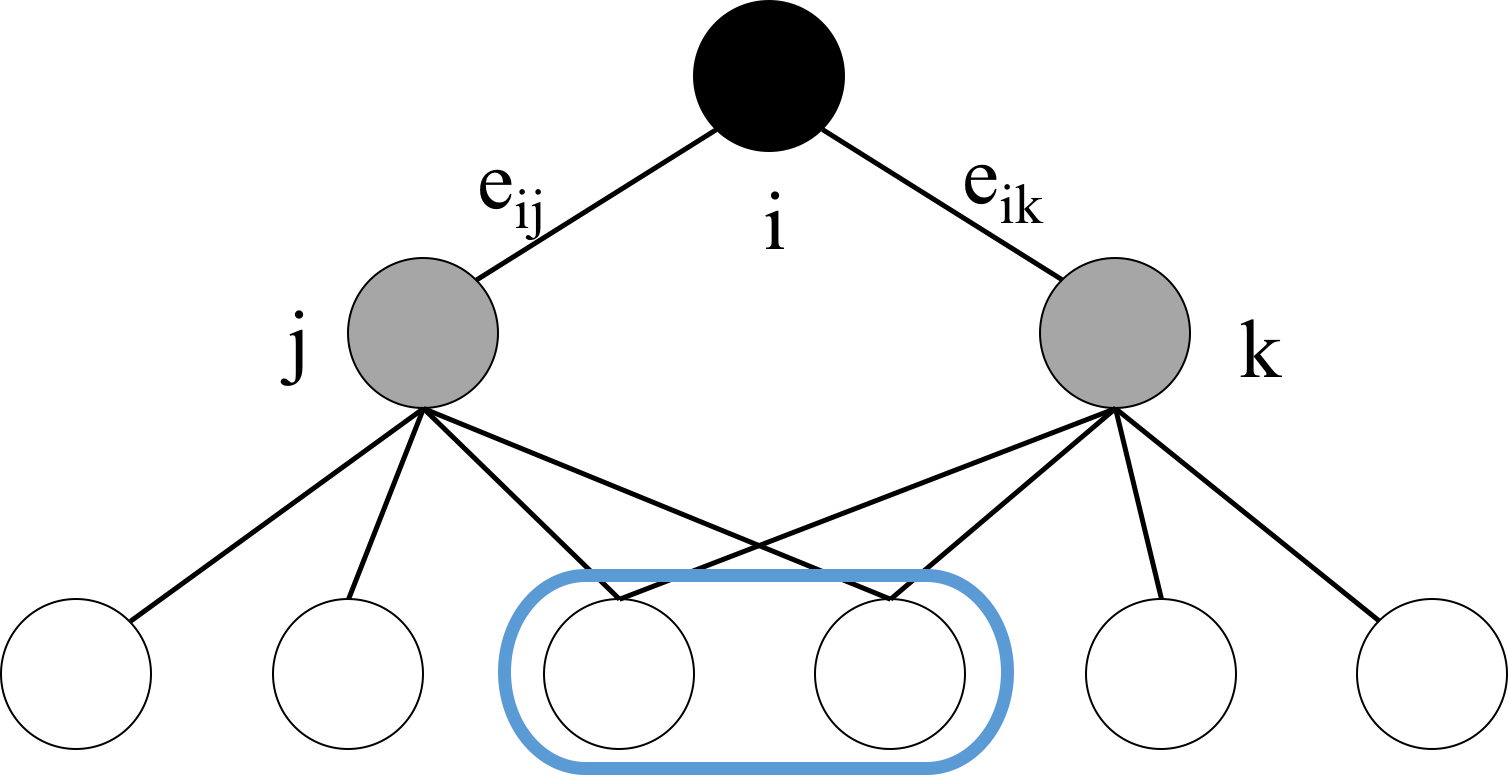}
\caption{Overlapped Nodes and Node similarity}
\label{node_similarity}
\end{figure}

In Figure~\ref{node_similarity}, we show the correlation of overlapped nodes and node similarity. In node-based community detection methods, the identification of communities is typically based on the connectivity patterns among nodes in a network. These methods aim to partition the network into cohesive groups or communities, where nodes within a community exhibit strong interconnectivity while having fewer connections to nodes outside the community.

One challenge that can arise in node-based community detection is the problem of overlapped nodes. Overlapped nodes refer to nodes that belong to multiple communities simultaneously, blurring the boundaries between communities. This means that these nodes have significant connections to nodes in multiple communities, making it difficult to assign them to a single community without sacrificing the accuracy of the community detection process.

The presence of overlapped nodes can introduce complications and ambiguity in community detection results. It can lead to difficulties in accurately identifying and delineating the boundaries of distinct communities within the network. Overlapping nodes can create overlaps between detected communities, causing them to merge or appear less cohesive than they actually are. This can impact the quality and meaningfulness of the community structure revealed by the detection method.

At first, we regard each edge as a community. After that, the edges will be enrolled into different communities. The enrollment criteria are edges share the same nodes with the first edge. The similarity of an edge pair $(e_{ij}),e_{ik}$ with a common node $u_i$ is to consider the similarity of $u_k$ and $u_j$. In the paper, we consider a concise but useful way, which is to evaluate the amounts of neighbors of $u_k$ and $u_j$. Based on this methodology, we formulate the similarity of $(e_{ij}),e_{ik}$ as

\begin{equation}
\label{e_edge_similarity}
\begin{aligned}
S\Big(e_{ij},e_{ik}\Big) = \frac{\Big| n_+(k) \cap n_+(j)\Big|}{\Big| n_+(k) \cup n_+(j) \Big|},
\end{aligned}		
\end{equation}

where $n_+(k)$ is the set of node $u_k$ and all its adjacent neighbours while $n_+(j)$ is the set of node $u_j$ and all its adjacent neighbours.

Through calculating the edge pairs' similarity, it is able to detect the SHN community by clustering in a hierarchical manner. First, all similarity values of possible edge pairs are calculated. Then, the similarity values are ordered descendingly. A tree graph structure is then established to merge communities in an iterated way. In the iteration process, if there exist edge pairs that share the same similarity, they shall be merged in the same round. The convergence of community merging is controlled by a threshold. Otherwise, all the edges are merged into one single community.

To better explain, the edge pair similarity can be regarded as the strength of the merged community. This also relates to the height of a branch of the tree graph structure. Therefore, to get reasonable communities, the key is to identify the best position to ``cut'' the tree, in particular, deriving the threshold of the community merging process. To avoid empirical errors, we establish an objective function called partitioning function $D(\cdot)$, based on the density of all possible edge pairs.

Let $M$ be the number of edge pairs inside a smart healthcare network, $|C|$ be the number of communities $\{P_1, P_2,..., P_c\}$, $m_c$ and $n_c$ are the number of edge pairs and nodes inside community $P_c$, the corresponding normalization density is

\begin{equation}
\label{e_partition_density}
\begin{aligned}
D_c = \frac{m_c - \Big(n_c - 1 \Big)}{n_c \Big(n_c -1\Big)/2 - \Big(n_c -1\Big)},
\end{aligned}		
\end{equation}

where $n_c - 1$ is the minimum number of edge pairs required to constitute a connected graph and $n_c(n_c-1)/2$ is maximum number of possible edge pairs among $n_c$ nodes. A special consideration is that $D_c = 0 $ if $n_c = 2$. Thus, the partition density of the whole network is formulated as the weighted sum of $D_c$.

\begin{equation}
\label{e_partition_density_whole}
\begin{aligned}
D = \frac{1}{M} \sum_c m_c D_c = \frac{2}{M} \sum_c m_c \frac{m_c - \Big(n_c-1\Big)}{\Big(n_c -2 \Big) \Big(n_c -1 \Big)}
\end{aligned}		
\end{equation}

From Eq.~\ref{e_partition_density_whole}, each term in the summation has a physical meaning within the community. Thus, the distinguishability limitation problem of modularity can be well mitigated. We can either directly optimize the partition density. The primary advantageous feature of this model is that we can flexible partition the community based on the requirements. Another advantage of using edge pairs tree graph is to reveal the hierarchy community structure feature by non-optimization cut-off rule.

\begin{figure}[!htbp]
\centering
\subfloat[]{
\includegraphics[width=1.4in]{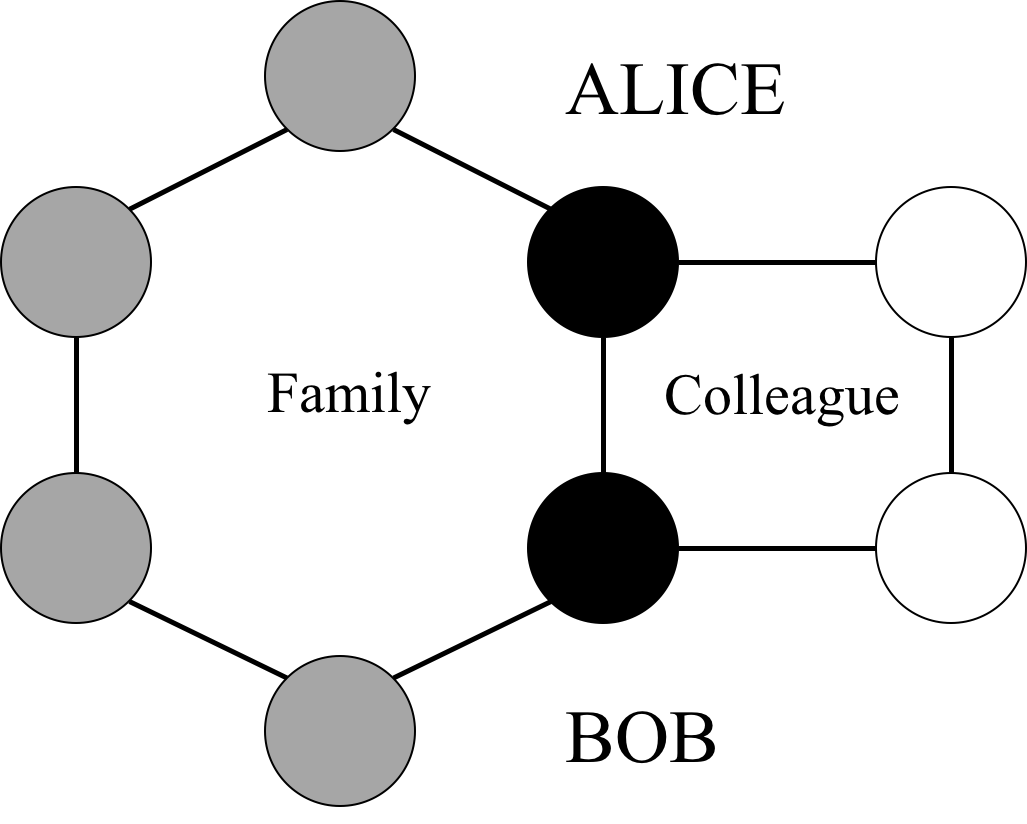}
\label{f_2a}
}
\hfil
\subfloat[]{
\includegraphics[width=1.4in]{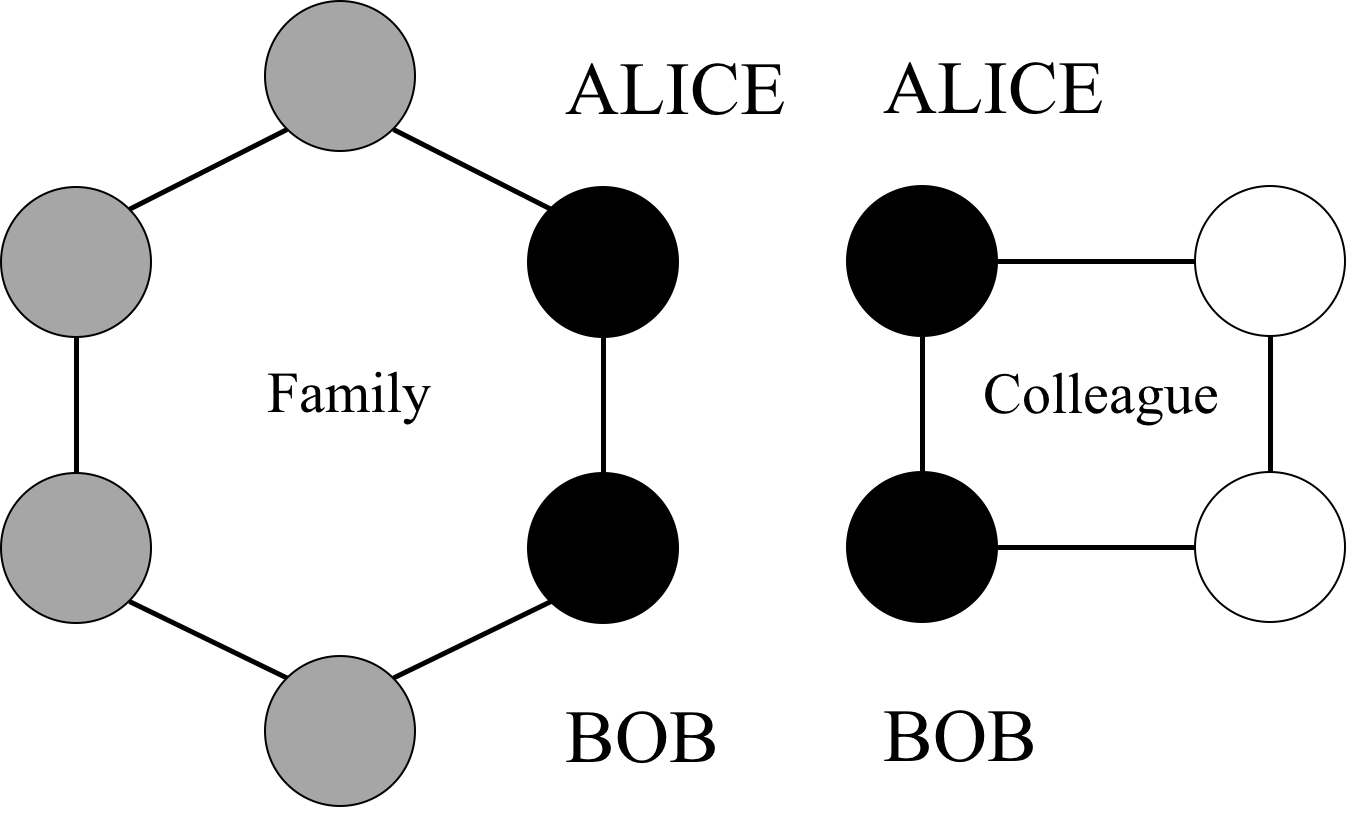}
\label{f_2b}
}
\caption{Edge Pair Link Community to Node Community}
\label{f_link_node}
\end{figure}

\subsection{Mapping Function Derivation}

To map the community density (trust) to a reasonable privacy protection level, a mapping function is required. The investigation shows that the sigmoid function is a good match in this scenario. As a function that is used to evaluate the quality of service (QoS), we modify it to fit the proposed model. Community density is used as the input to generate personalized privacy protection levels, namely, the value of $\epsilon$. The mapping function is referred to as QoS-based mapping function in this context.

The partitioning mechanism described above is used to generate several communities $C = \{P_1, P_2, ..., P_c\}$. Then we calculate the density of each community and further derive the personalized $\epsilon$ values. Apparently, the number and density of community may correspond to various users in SHN. The $\epsilon$ value should not linearly increase with the density as well. Thus, the sigmoid function works well in this scenario.

The reason why a sigmoid function is chosen is due to its unique features. To begin with, when the community density $D_c$ is small (e.g., the most special case is a community formed by another single user), the privacy level should be high, and it increases slowly with the increase of the density. After the community has a certain scale $D_c$, the privacy level can be relaxed, and the relaxing rate is relatively high. However, when the community density $D_c$ is large enough, the community is trusted in a high possibility, and the privacy level is relatively low. The impact of further relaxation is marginal, and thereby the increased scope slows down. 

The modified mapping function is defined as

\begin{equation}
\label{e_sigmoid_function}
\begin{aligned}
\epsilon_i = f(i) = \omega \times \frac{1}{1 + \exp(-\theta / D_c - \alpha)},
\end{aligned}		
\end{equation}

where $\omega$ is the weight parameter to adjust the amplitude of the maximum value, $\theta$ is leveraged to decide the steepness of the curve, and $\alpha$ denotes the location of the symmetric line.

Moving on, after Equ.~\ref{e_partition_density} is substituted into Equ.~\ref{e_sigmoid_function}, the Sigmoid function is reshaped into

\begin{equation}
\label{e_sigmoid_function_2}
\begin{aligned}
\epsilon_i = f(i) = \omega \times \frac{1}{1 + \exp \Big(-\theta \cdot \frac{n_i (n_i -1)/2 - (n_i -1)}{m_i - (n_i - 1 )} - \alpha \Big)},
\end{aligned}		
\end{equation}


\subsection{Community Density-based Personalized Privacy Protection in Smart Healthcare Networks}

Users who share the same interest or experience similar symptoms usually join the same community in a SHN. Besides, there is high possibility for them to form similar communities in other SHNs. Since the shared data over SHNs are highly sensitive, it is strictly necessary that other users can only access the processed data constrained by certain privacy protection methods.

\begin{figure}[!htbp]
\centering
\includegraphics[width=4.2in]{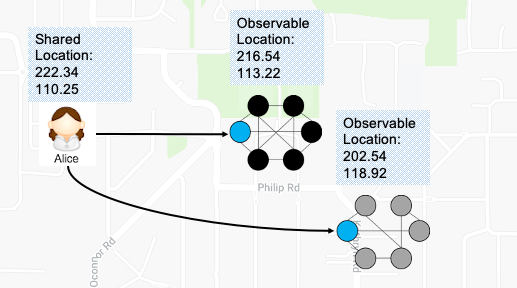}
\caption{Personalized Privacy Protection Instance}
\label{f_location}
\end{figure}

The shared data is usually with various auxiliary information like location. As mentioned above, the community with higher density receives more accurate data while the low-density community receives less accurate data, which is shown in Fig.~\ref{f_location}. Alice belongs to two different communities. After executing Algorithm~1, personalized privacy protection levels are derived using Eq.~\ref{e_sigmoid_function_2}. After that, the corresponding location information is shared to two communities with different accuracy based on different privacy protection levels. However, the shared data in different communities may be used to launch linkage attacks, which is discussed in the following subsections.

Theoretically, users that are more trustworthy will receive more accurate data. On the contract, users that are less trustworthy will receive less accurate data. To evaluate the trust among users, we use a simple and straightforward index, which is community density. Other indexes may be applicable in other scenarios. If the density is large, then people share a higher level of trust and vice versa. Then, a lower level of privacy protection will be acted on the raw data and people within this community can access reasonably accurate data.

Built upon the classic differential privacy, we formulate the personalized differential privacy as follows.

\begin{myDef}[Personalized $\epsilon$-Differential Privacy]
\label{def_differential_privacy}

Given $\epsilon \geq 0$, $D$ to be the space of the sensitive data, and $\mathcal{A} \subseteq D \times D$ to denote an adjacent relation. A mechanism is $\mathcal{M} \to \Delta(\mathcal{Y})$considered to be $\epsilon$-differentially private if
\begin{equation}
\label{e_differential_privacy}
\begin{aligned}
\Pr\Big[\mathcal{M}(D) & \in \Omega\Big] = \exp(\epsilon) \cdot \Pr\Big[\mathcal{M}(D') \in \Omega\Big],\\
\end{aligned}		
\end{equation}
where $\mathcal{Y}$ is the noisy outcome and $\epsilon$ is the privacy protection level that varies with the community density.

\end{myDef}

The privacy protection level $\epsilon$ is defined in Equ.~\ref{e_sigmoid_function_2}. Therefore, if we substitute it into Equ.~\ref{e_differential_privacy}, we have 

\begin{equation}
\label{e_differential_privacy_2}
\begin{aligned}
\frac{\Pr\Big[\mathcal{M}(D) \in \Omega\Big]}{\Pr\Big[\mathcal{M}(D') \in \Omega\Big]} = \exp \Bigg(\frac{\omega \exp \Big(\theta \cdot \frac{n_i (n_i -1)/2 - (n_i -1)}{m_i - (n_i - 1 )} + \alpha \Big)}{1 + (1-\alpha)\exp \Big(-\theta \cdot \frac{n_i (n_i -1)/2 - (n_i -1)}{m_i - (n_i - 1 )} \Big)} \Bigg), 
\end{aligned}	
\end{equation}
where the conditions of Equ.~\ref{e_differential_privacy_2} inherits from Equ.~\ref{e_differential_privacy}. In the proposed personalized $\epsilon$-differential privacy model, the privacy level $\epsilon$ is personalized by the density $D_c$ through a sigmoid function.

In the personalized differential privacy model, $\epsilon$ is an index to evaluate the privacy protection level and data utility since it decides the volume of the randomized noise. The overall privacy budget $B$ is the total value of all possible $\epsilon$. To measure the data utility, one of the most popular matrices in this scenario, root-mean-square-error (RMSE) is used to describe the balance between privacy protection and data utility.\footnote{The data utility denotes how much useful information is left in the sanitized data. It is an important parameter to measure the effectiveness of privacy protection models.}

\begin{equation}
\label{e_rmse}
\begin{aligned}
RMSE = \sqrt{\sum^n_{i=1}\sum^n_{j\neq i} E\Big|\Big|y_{ij}-d_i\Big|\Big|^2_2}
\end{aligned}		
\end{equation}

From the perspective of the optimized tradeoff between privacy protection and data utility, given overall privacy budget $B$ and minimum data utility $\min(RMSE)$, we have

\begin{equation}
\label{e_tradeoff}
\begin{aligned}
\mathrm{Optimize} & \ \mathrm{Tradeoff:} \  \max(\epsilon), \max(RMSE)   \\
s.t. & \\
&\epsilon_i =\frac{\omega}{1 + \exp \Big(-\theta \cdot \frac{n_i (n_i -1)/2 - (n_i -1)}{m_i - (n_i - 1 )} - \alpha \Big)}, \\
& \sum_i^n \epsilon_i \geq B, \\
& \sqrt{\sum^n_{i=1}\sum^n_{j\neq i} E\Big|\Big|y_{ij}-d_i\Big|\Big|^2_2} \geq \min(RMSE). \\
\end{aligned}		
\end{equation}

\subsection{Linkage Attack Model}

In this subsection, we establish the adversary model and attack model. To qualitatively evaluate the adversaries and attacks, we model them using differential privacy mathematical theories. 

For adversaries, we practically assume that they have a certain amount of background knowledge. The amount of background knowledge can be adjusted. The linkage attack is launched by such adversaries, who also has the access to multiple data sources. In SHNs, one of the most useful and easy-to-access background knowledge is the user connections (a sub-graph in the system model).

In existing research, the modeling of the background knowledge of the adversary and linkage attack is barely discussed. As discussed above, the background knowledge can be regarded as raw data with noise, which is the same as differentially private noise in nature. Therefore, the linkage attack is the aggregation of mutliple pieces of raw data with different noises. Built upon this assumption, linkage attack is formulated as

\begin{myDef}[\textbf{Linkage Attack}]
\label{def_linkage_attack}

Given multiple data resources $\{D_i | i = 1, 2, 3, ..., n\}$, a linkage attack $L(\cdot)$, and $\mathcal{M}$ to be a randomized algorithm that sanitizes the dataset where $\epsilon_i = \mathcal{M}(D_i)$, the linkage attack is successfully launched if 

\begin{equation}
\label{e_linkage_attack}
\begin{aligned}
\sum_i^n \mathcal{M}(D_i) \geq L(\cdot) \geq \max \Big\{\mathcal{M}(D_i)\Big| i = 1,2,...,n \Big\}\\
\end{aligned}		
\end{equation}

\end{myDef}

Under this assumption, the success criteria of a linkage attack can be expressed as a release of differential privacy protection level. Quantitatively, the value of $\epsilon$ will increase as the attack result. To maximize the performance, we consider the worst-case linkage attack and corresponding countermeasure, noise decoupling mechanism as follows.

\subsection{Noise-Decoupling Mechanism}

We start with two randomized algorithms, namely the Laplace mechanism and the exponential mechanism. Then, the noise-decoupling mechanism is given to improve the performances of the two algorithms.

\subsubsection{Laplace Mechanism and Exponential Mechanism}

As mentioned above, $\epsilon$-differential privacy is a probabilistic definition. It is necessary to design some randomized mechanisms which are differentially private. In this subsection, we will introduce two of the randomized algorithms, which are the Laplace mechanism and exponential mechanism, respectively.

In numeral scenarios, the Laplace mechanism is widely used to inject random noise under certain control. The noise generation compiles with a Laplace mechanism as

\begin{equation}
\label{e_laplace_mechanism}
\begin{aligned}
Lap\Big(\frac{\delta}{\epsilon}\Big) \sim d\Pr\Big[M=m\Big] = \exp\Bigg(-\frac{||m||_2 \times \epsilon}{\delta}\Bigg),
\end{aligned}		
\end{equation}

where $e^{\frac{|f(D_1)-f(D_2)|}{\lambda}} \leq e^{\delta(f)}$. We regard the $\delta(f)$ as privacy level $\epsilon$. Although Gaussian noise can also be utilized to achieve differential privacy, it requires a slight relaxation of the definition of differential privacy. Therefore, we employ Laplace noise to deal with the numeral data published by users.

Besides the Laplace mechanism, we need to introduce an exponential mechanism to the proposed model as well. The reason is the Laplace mechanism is limited in the scenario of numeral data. However, the exponential mechanism functions well to handle textual data.

Let $O$ be the set of candidate items, $o_i \in O$ be a candidate item, $f(D,o)$ be the function that outputs the number of $o_i$ inside $D$, a mechanism $M_f^{\epsilon}(D)$ is $\epsilon$-differentially private if the probability of $o_i$ being the output   is proportion to $e^{\frac{\epsilon f(D,o)}{2\delta f}}$.

\begin{equation}
\label{e_exponential_mechanism}
\begin{aligned}
\frac{Pr[M_f^{\epsilon}(D) = o]}{Pr[M_f^{\epsilon}(D') = o]} = \Bigg( \frac{\exp\Big( \frac{\epsilon f(D,o)}{2 \delta f} \Big)}{\exp\Big( \frac{\epsilon f(D',o)}{2 \delta f} \Big)} \Bigg) \Bigg( \frac{\sum_{o'} \exp\Big( \frac{\epsilon f(D',o')}{2 \delta f} \Big)}{\sum_{o'} \exp\Big( \frac{\epsilon f(D,o')}{2 \delta f} \Big)} \Bigg) \leq \exp(\epsilon)
\end{aligned}		
\end{equation}

In order to control the noise value, we introduce another term,  which is global sensitivity. Assume a function $f: D \to R^d$, we have an input dataset, and the output should be a $d$ dimensional real-valued vector. For any adjacent dataset $D$ and $D'$, the global sensitivity is defined as

\begin{equation}
\label{e_global_sensitivity}
\begin{aligned}
G_f = \max_{D,D'} \Big|\Big|f(D) -f(D') \Big|\Big|.
\end{aligned}		
\end{equation}

The global sensitivity can apply to both the Laplace mechanism and exponential mechanism and helps to determine privacy and accuracy.

\subsubsection{Decoupling the Correlation among Noises}

For $n$-tuple real-valued data $d$, we aim to propose a private mechanism $\mathcal{M}$ to generate the approximation $y_{ij}$, in which $y_{ij}$ is sent from $u_i$ to $u_j$. As shown in Algorithm~2, two features are required for the mechanism $\mathcal{M}$. Firstly, the absolute error $|y_{ij} - d_{ij}|$ is supposed only to be determined by the density of the community. The rest of the arguments should have no impact on the absolute error. Secondly, the linkage of any series of data resources will not reveal further sensitive information about the individual. The workflow of the proposed attack-proof personalized differential privacy is shown by the pseudo-code in Algorithm 2.

\renewcommand{\algorithmicrequire}{\textbf{Input:}} 
\renewcommand{\algorithmicensure}{\textbf{Output:}}
\begin{algorithm}[!h]  
    \label{algorithm2}
    \caption{Attack-Proof Personalized Differential Privacy}
    \begin{algorithmic}[1]
    \REQUIRE Raw data $D$;
    \ENSURE Sanitized data set $D_i'$ with personalized privacy protection;
    	\STATE Derive communities and community densities $D_c$;
	\STATE Set up proper sigmoid function $\frac{1}{e^{x} + 1}$; 
	\IF {$u_i$ belongs to a single community $P_c$}
	\STATE Personalize privacy level $\epsilon_i$ based on $D_c$;
	\ELSE
	\STATE Personalize privacy level $\epsilon_i$ based on $\min(D_c)$;
	\ENDIF
	\IF {Data $D$ is numeral data}
	\STATE Choose Laplace mechanism;
	\ELSE
	\STATE Choose exponential mechanism;
	\ENDIF
	\STATE Deploy noise decouple mechanism;
	\STATE Deploy data utility optimization mechanism;
        \STATE Release $D_i'$ to different communities $C = \{P_1, P_2, ..., P_c\}$.
    \end{algorithmic}  
\end{algorithm}

Algorithm~2 takes raw data, D, as input and aims to generate a sanitized data set, D', with personalized privacy protection. It begins by deriving communities and computing community densities, Dc, from the raw data. A proper sigmoid function is set up to calculate personalized privacy levels for individual data points. The algorithm then determines the privacy level, $\epsilon_i$, based on whether a data point, $u_i$, belongs to a single community or multiple communities. For numerical data, the Laplace mechanism is chosen to add privacy-preserving noise, while for non-numerical data, the exponential mechanism is used. The algorithm employs noise decoupling and data utility optimization mechanisms to enhance privacy and maintain data quality. Finally, the sanitized data, D'i, is released to different communities, C = {P1, P2, ..., Pc}, considering the personalized privacy levels and community memberships of the data points. The output is a privacy-protected and community-tailored sanitized data set, D'.

The algorithm starts by deriving communities and community densities from the raw data D. The time complexity of this step depends on the specific community detection method employed and can range from O($n^2$) to O($n^3$), where n represents the size of the raw data. Community detection often involves iterative processes, graph traversal, or optimization algorithms, contributing to the complexity.

Next, the algorithm sets up a sigmoid function, which has a constant time complexity of O(1) as it involves defining a mathematical function. The algorithm then checks whether a data point, ui, belongs to a single community or multiple communities. This step has a constant time complexity of O(1) as it involves a simple conditional check.

Based on the membership status, the algorithm personalizes the privacy level, $\epsilon_i$, for the data point. The time complexity of this step depends on the computation involved in assigning a personalized privacy level based on the community densities, Dc. It can range from O(1) to O($|C|$), where $|C|$ represents the number of communities.

The algorithm proceeds to choose either the Laplace or exponential mechanism for privacy preservation. This step has a constant time complexity of O(1) as it involves selecting between two predefined mechanisms.

Next, the algorithm deploys the noise decouple mechanism to protect privacy. The time complexity of this step depends on the specific method used for noise decoupling and can range from O(n) to O($n^2$), where n represents the size of the data. Complex computations or iterative processes may be involved in this step.

Similarly, the algorithm deploys the data utility optimization mechanism, which aims to balance privacy and data quality. The time complexity of this step depends on the specific optimization methods used and can range from O(n) to O($n^2$), depending on the size of the data and the complexity of the optimization process.

Finally, the algorithm releases the sanitized and privacy-protected data, D'i, to different communities. This step has a constant time complexity of O(1) as it involves the release of data to predefined communities.

In summary, the overall time complexity of the algorithm depends on the specific methods used for community detection, privacy level personalization, noise decoupling, and data utility optimization. The time complexity ranges from O($n^2$) to O($n^3$) for community detection, and the other steps generally have time complexities ranging from O(1) to O($n^2$), depending on the size of the data and the specific computations involved in each step.

Observe in Algorithm~2 that to achieve the targets, we generate the noises defined on a private stochastic process, which is designed and discussed in detail as below.

\begin{myDef}[\textit{\textbf{Noise-Decoupling Process}}]
\label{def_private_stochastic_process}

Let $\epsilon$ be the privacy budget (defined by differential privacy) and $\epsilon_i$, $\epsilon_{i+1}$,$\epsilon_{i+2}$ be three instances of privacy protection level. Given $\epsilon_i < \epsilon_{i+1}<\epsilon_{i+2}$, the following properties are required for the private stochastic process.

\begin{itemize}

\item[$\circ$] The noise complies with Laplacian Mechanism: $\forall \epsilon > 0, d\Pr\Big(V_{\epsilon} = v\Big) \propto \exp\Big(-\epsilon ||v ||_2 \Big)$;

\item[$\circ$] The noise generation process complies with the private stochastic process: $\forall \epsilon_i < \epsilon_{i+1}< \epsilon_{i+2}, V_{\epsilon_i} | V_{\epsilon_{i+1}}, V_{\epsilon_i} \bot V_{\epsilon_{i+2}}$;

\item[$\circ$] The transfer probability in the Markov process is
\begin{equation}
\label{e_transfer_probablity}
\begin{aligned}
& d\Pr\Big(V_{\epsilon_i}=v_i \Big| V_{\epsilon_{i+1}}=v_{i+1} \Big) \propto \delta (v_i-v{i+1}) \\
& + \frac{(n+1)\epsilon_i^{1+\frac{n}{2}}||v_i-v_{i+1}||_2^{1-\frac{n}{2}}}{(2\pi)^{\frac{n}{2}}}B_{{\frac{n}{2}}-1}\Big(\epsilon_i ||v_i-v_{i+1}||_2 \Big)\tau + O\Big(\tau^2 \Big), \\
							   & \textbf{s.t.}  \\
							   &  \quad  \quad \quad \tau = \frac{\epsilon_i}{\epsilon_{i+1}} - 1,  \\		
\end{aligned}
\end{equation}
where $B$ is a Bessel function.

\end{itemize}

\end{myDef}

\section{Blockchain-Enhanced Mechanism Against Data Falsification}
\label{sec::attack_proof}

In this part, we present how the tailor-made blockchain structure can guarantee the integrity of differentially private health data and prevent data falsification operations.

\subsection{Consortium Blockchain-based Smart Healthcare Network}

\begin{figure}[!htbp]
\centering
\includegraphics[width=3.2in]{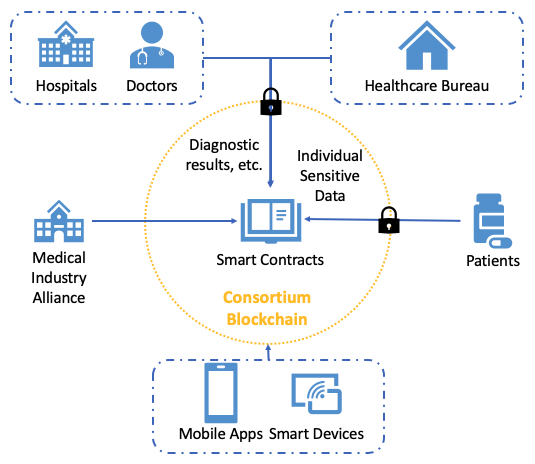}
\caption{Consortium Blockchain based Smart Healthcare Network}
\label{f_blockchain}
\end{figure}

The smart healthcare network has been modeled as a graph structure as above. For each node in the graph, it is also a node in the proposed consortium blockchain system as well, as shown in Fig.~\ref{f_blockchain}. From the figure, we can tell there are different entities in SHNs, including but not limited to patients, hospitals, health bureaus, etc. Different entities have different accesses. For example, the health bureau has the most access due to its supervision role, but it does not have the right to revise the data stored on-chain. Hospitals have access to the health history of patients and are allowed to add new items to the history. Patients can only access their own data with the rights of adding, deleting, or revising \cite{qu2023learn}. Except for this, all parties conduct data sharing which dynamically changes the access during the operation of the system

As mentioned earlier, doctors and patients can manage the data and thereby will generate blocks of the consortium blockchain. Usually, the block will include some sensitive information like diagnosis notes, identity information, time stamp, location, etc. The privacy protection techniques of the data stored on the consortium blockchain are a necessity and will be discussed in the subsequent subsections.

To make it more secure and robust, the Proof-of-Work (PoW) consensus algorithm is deployed. This requires miners to mine for a nonce value and get the block generation chance. The data is firstly broadcast to all eligible parties and all parties start to calculate the nonce value. The party that first finds the nonce will generate a candidate block, after which, the block is broadcast to all eligible parties again. The parties who receive a candidate block of this round stops mining and validates the data in the block. If the data is authentic, the block will be appended to the local chain. As a consortium blockchain, the health bureaus will serve as the leader of the chain to help with the consensus process. We can always add more layers for more functionalities. For example, an analytic layer could be used for disease surveillance \cite{DBLP:journals/tcss/WangWWQYOGW18}. In such a consortium of blockchain-based smart healthcare networks, several privacy requirements are to be met, including data integrity, and interoperability, specifically for healthcare research facilitation.

\subsection{Blockchain Brief Overview}

We show the basic PoW-based blockchain structure in this subsection, which is the foundation of the proposed structure.

\begin{figure}[!htbp]
\centering
\includegraphics[width=4.2in]{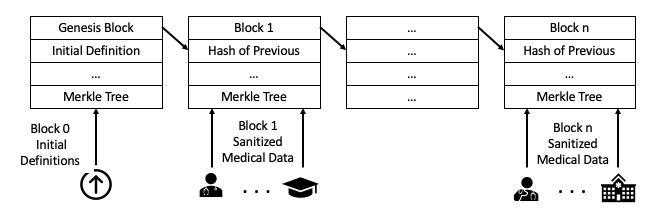}
\caption{Blockchain Architecture for Sanitized Sensitive Medical Data Sharing}
\label{f_medical}
\end{figure}

In Fig. \ref{f_medical}, a generalized structure of blockchain systems for medical data sharing in smart healthcare networks is presented. This structure demonstrates how blocks are appended to each other, ensuring a sequential order of transactions. Each block includes the hash value of the previous block, creating a chain-like structure that ensures the integrity and immutability of the data.

Within each block, transactions are stored in a Merkle tree structure. The Merkle tree allows for efficient summarization and verification of the transactions contained within the block. By hashing the individual transactions and then combining them in pairs until a final hash value, known as the Merkle tree root, is computed, the block can represent a condensed representation of all the transactions it contains.

In this particular context, a transaction refers to a piece of sanitized medical data that adheres to personalized differential privacy. Sanitization techniques are applied to the medical data to remove personally identifiable information and ensure individual privacy. Personalized differential privacy ensures that the level of privacy protection is tailored to the specific requirements and preferences of each patient or data subject.

By utilizing blockchain technology in the sharing of medical data, the structure depicted in Fig. \ref{f_medical} offers several benefits. It enhances data security, as the hash values and the chaining mechanism make it extremely difficult for unauthorized parties to tamper with or modify the stored data. The Merkle tree structure facilitates efficient and secure verification of the integrity of transactions within a block. Additionally, the use of personalized differential privacy techniques helps protect the privacy of individuals while enabling the sharing and analysis of aggregated and anonymized medical data.

Overall, the blockchain system presented in Fig. \ref{f_medical} provides a robust framework for secure and privacy-preserving medical data sharing in smart healthcare networks, ensuring data integrity, tamper resistance, and individual privacy protection.

In Table. \ref{tab:tableX}, an instance of a generalized block header is shown with two key components. The block header usually contains a version of the block, parent block hash, Merkle tree, timestamp, nBits, and Nonce (for Proof-of-Work). The size of the block determines the maximum number of transactions (shared data). The shared data is usually protected and validated by asymmetric cryptography in a trustworthy case. But if there is an untrustworthy environment, other techniques may be deployed, such as a digital signature. In the block body, it contains sanitized heath data or any other data types regarding the application scenarios.

\begin{table}[!t]
\centering
\caption{An Instance of a Block Header in the Blockchain System}
\label{tab:tableX}
  \begin{tabular}{ c c}
    \hline
    \hline
    Element  & Example Value     \\
    \hline
    Block Version & 03000000 \\
    \hline
    Parent Block Hash  & \tabincell{c}{c5ee0a2b1480a2852a30d5ffe364d98e\\10d9334beb48ca0d000000000000000} \\
    \hline
    Merkle Tree Root & \tabincell{c}{8c2de567cad23df7992e030b44af454d \\70add80201edcd21cbb940ab88da45c}\\
    \hline 
    Timestamp  & 29d5a5a4 \\
    \hline 
    nBits  & 34cd1b29 \\
    \hline 
    Nonce  & fd8e2664 \\
    \hline
    \hline
  \end{tabular}
\end{table}

\begin{itemize}
    \item[$\circ$] \textbf{Version of block:} The block version is a numeric value that represents the version of the blockchain protocol being used. It helps ensure compatibility between different versions of the blockchain software. The version number may be updated over time as new features or improvements are introduced to the protocol.
    \item[$\circ$] \textbf{Parent block hash:} The parent block hash is a unique identifier for the previous block in the blockchain. It is the hash value of the header of the previous block. By including the parent block hash in the current block header, the blockchain maintains a chronological order and creates a chain of blocks. This chaining mechanism ensures the immutability and integrity of the blockchain.
    \item[$\circ$] \textbf{Merkle tree:} The Merkle tree root is a hash value that represents a condensed summary of all the transactions within the block. In a blockchain, transactions are grouped together in a Merkle tree structure. The Merkle tree allows for efficient verification of the integrity of all the transactions in the block. The root hash is calculated by hashing the concatenated hash values of the individual transactions in a specific order until a single hash value remains.
    \item[$\circ$] \textbf{Timestamp:} The timestamp indicates the time when the block was created or mined. It is typically represented as a Unix timestamp, which is a numerical value that represents the number of seconds elapsed since January 1, 1970. The timestamp helps establish the order of blocks in the blockchain and ensures that blocks are added at regular intervals.
    \item[$\circ$] \textbf{nBits:} The nBits value represents the target difficulty for mining the block. Mining is the process of finding a nonce value that, when combined with other block data, produces a hash value below a certain target difficulty. The nBits value encodes the target difficulty, which determines the computational effort required to mine a block. Miners adjust the nBits value periodically to maintain a consistent block generation rate.
    \item[$\circ$] \textbf{Nonce:} The nonce is a random value that miners modify during the mining process in order to find a suitable hash value that satisfies the target difficulty. Miners repeatedly change the nonce and recompute the block hash until they find a nonce that, when combined with other block data, produces a hash value that is below the target difficulty. The nonce is a crucial component of the proof-of-work consensus algorithm, which ensures that mining requires computational effort and contributes to the security of the blockchain.
\end{itemize}

These elements, combined together, form the block header. The block header is hashed to produce the block's unique identifier, which is used in the blockchain's consensus algorithm to validate and add the block to the blockchain.

The connection between blocks ensures the integrity and immutability of the blockchain. Any change to a block's data would alter its hash, making it inconsistent with the stored parent block hash in the subsequent block. This would break the chain and invalidate the affected block, alerting the network to potential tampering attempts. Therefore, altering the data in one block would require recalculating the hash for that block and all subsequent blocks, which becomes increasingly computationally expensive and practically infeasible as the blockchain grows longer.

This linking mechanism provides several important benefits. \textbf{Security}: The chain of blocks ensures the security of the blockchain by preventing unauthorized modifications. Once a block is added to the blockchain, it becomes extremely difficult to alter any past blocks without the consensus of the majority of network participants. \textbf{Immutability}: The chaining of blocks creates an immutable record of transactions. Once a block is added to the blockchain, its contents are effectively set in stone. This feature is valuable in applications where data integrity and auditability are critical. \textbf{Consensus}: The connection between blocks plays a crucial role in achieving consensus among network participants. By following the longest valid chain of blocks, participants can agree on the order of transactions and determine the valid state of the blockchain.

\subsection{Importance of Defense against Data Falsification}

In all data-sharing scenarios or services, the data falsification issue is an inescapable topic due to its significant negative impact. For instance, one of the primary attacks in this domain, poisoning attacks, is usually launched by adversaries. To poison the raw data, adversaries may inject or replace the data with misleading features. This can bring catastrophe to smart healthcare applications since the data are directly related to human health. To mitigate the risks, we explain the devised approach in the following paragraphs.

On the tailor-made blockchain, the differentially private data (raw data + differentially private noise) are saved with certain access control. When an individual queries the blockchain, the differentially private data will directly used as the response to the querier. However, since access control is deployed, an individual can only access the data within the community but an individual can belongs to multiple communities as described above. Within a community, all members share a same privacy protection level and can access the same data. Different from traditional blockchain systems, it is partitioned into many consortium sub-chains, with which community members only need to maintain their corresponding sub-chain. This brings more efficiency and flexibility for smart healthcare network users.

Despite this, adversaries will still try to falsify the differentially private data stored on the sub-chains. The adversary may inject or revise the data to gain financial benefits. However, all behaviors will be cross-validated by community members of each sub-chain. In this case, if there is a malicious operation on data, most trustworthy members will choose not to act on it and thereby the operation cannot be performed. The raw data will remain unchanged in this case. At the same time, smart healthcare data sharing will not be as frequent as transaction systems, which will not bring in too much computation burden or processing delay. The sub-chain structure also can improve the efficiency.

\section{Performance Evaluation}
\label{sec:experiment}

This section will illustrate the experimental settings and results to validate the effectiveness of the proposed solution. The proposed model is abbreviated to C-DP while the two benchmark models are classic differential privacy (DP) and personalized differential privacy (P-DP), respectively. P-DP uses the virtual online distance as the personalization index. To evaluate the whole system, we evaluate privacy protection performance, data utility degree, community distribution similarity, and blockchain performance.

We use two real-world datasets for the experiments, which are \textit{Doximity} dataset and \textit{HealthTap} dataset \cite{doximity, healthtap}. \textit{Doximity} dataset is collected from \textit{Doximity} from its Developer API following a uniform distribution \cite{doximity}. \textit{Doximity} is a popular health social network that offers abundant functions like befriending, news publishing, data sharing, etc. The details of the obtained data are as follows. Doximity is one of the very few popular social networks for doctors that provides developer API, but it does not allow broad querying of its database. Therefore, we randomly obtained $2,122$ nodes and $14,389$ edges with an average degree of $4.39$. The collected data has been anonymized by \textit{Doximity}, but the nodes and edges relationship are stored for research purposes. In addition, we also built an anonymous doctor social graph via the API of \textit{HealthTap} \cite{healthtap}. Health information of the HealthTap is provided interactively by a network of nearly over $150,000$ licensed doctors. It also provides the functionality of peer review, which includes doctors rating each other and self-identifying specializations. From the HealthTap, we obtain a total of $1,325$ nodes consisting of a network of $5,231$ edges. 

\begin{figure*}[!h]
\centering
\subfloat[]{
\includegraphics[width=2.25in]{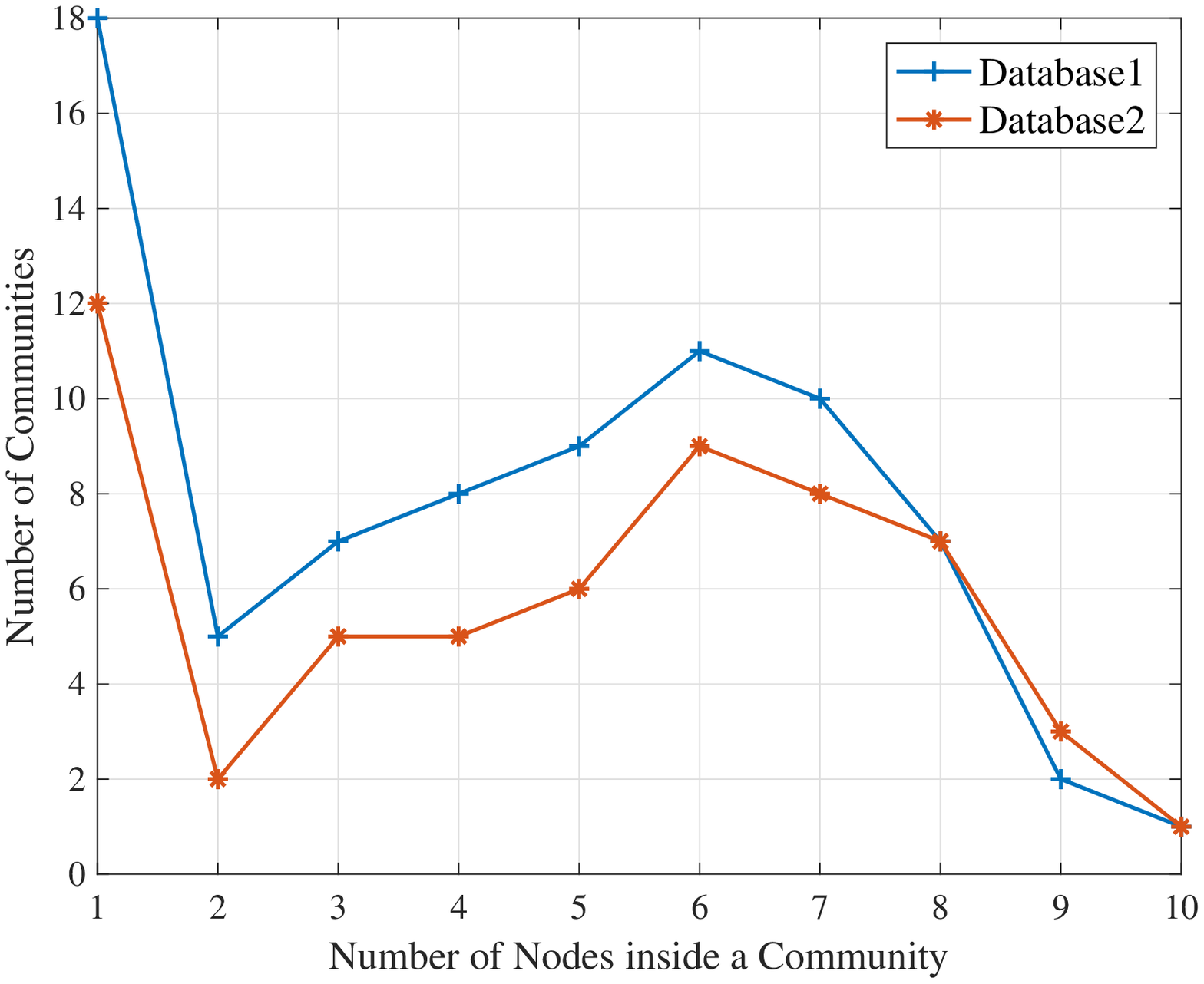}
\label{f_0a}
}
\hfill
\subfloat[]{
\includegraphics[width=2.25in]{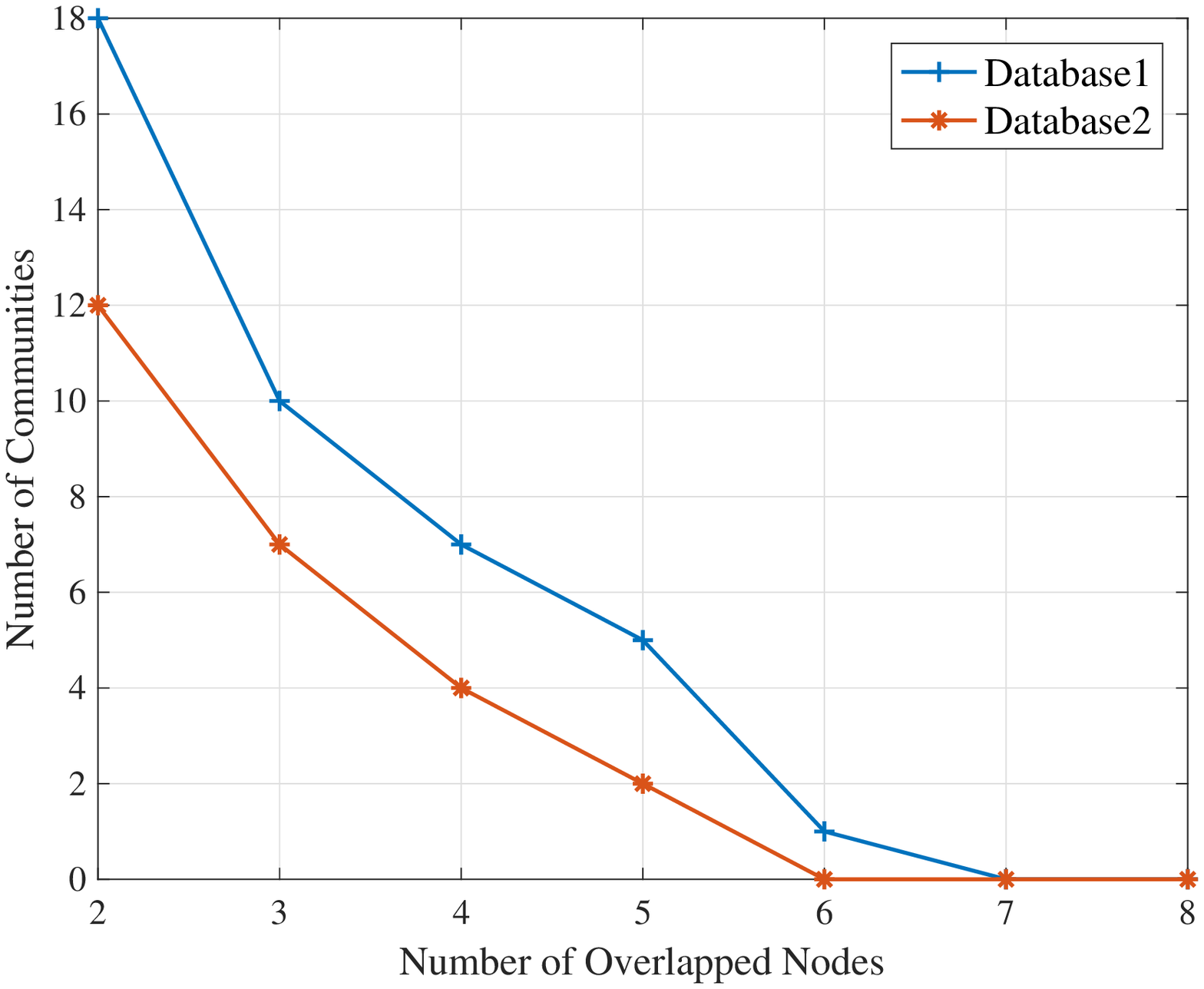}
\label{f_0b}
}
\caption{Community Similarity Distribution in the Smart Healthcare Network}
\label{figure1}
\end{figure*}

In this work, when we talk about DP, it is the classic differential privacy that provides uniform privacy protection to all users. In the case of P-DP, it is personalized privacy that leverages virtual online distance to personalize privacy levels. For C-DP, which is the proposed model, it maps community density to personalized privacy protection levels.

\subsection{Community Density Similarity}

In order to utilize community density as the index, we firstly use evaluation results to demonstrate the similar distribution of the two datasets. The outcome shows that both the community number distribution and the overlapped nodes distribution are quite close, which verifies our idea of mapping density to privacy level.

In Fig.~\ref{f_0a}, we compare the density with the node number. The distributions are quite similar to each other. Of all the records, there is a noisy point where node number equals $1$. When node number is $1$,  we can easily tell that they have no mutual trust people in this smart healthcare network. The amount of this node is normally large, which captures the real-world features.

In Fig.~\ref{f_0b},  we illustrate the distributions of overlapped nodes. Both of the numbers are decreasing with the number of overlapped communities increasing. They share the same trends and similar numbers while the maximum difference is no larger than $2$. The x-axis starts from $2$ since one node in one community is not defined as overlapping. In Fig.~\ref{f_0b}, we have tried to discuss one advanced feature with these two data sets, which shows the total amount of each group regarding the inclusion of overlapping nodes. Intuitively, any node can be assigned to many groups. Therefore, we show that a negative correlation between these two parameters. Besides, for the given datasets, we can observe that the existence of the community is questionable when the number of  nodes is $>6$.

The comparison of density with the node number (representing the number of nodes in a community) is shown. The distributions of density and node number are quite similar, indicating that communities in the smart healthcare network tend to have consistent densities regardless of their size. However, there is a notable noisy point where the node number equals 1. This implies that there are instances where a community consists of only one node, indicating a lack of mutual trust connections within that community. The occurrence of such single-node communities is relatively common, which aligns with real-world characteristics of smart healthcare networks.

The distributions of overlapped nodes are illustrated. The numbers of overlapped nodes decrease as the number of overlapping communities increases. This trend suggests that as nodes are assigned to more communities, the likelihood of them belonging to multiple communities decreases. The distributions of the two numbers (overlapped nodes and the number of overlapping communities) show similar patterns, with a maximum difference of no more than 2. This indicates a consistent relationship between the two parameters. Additionally, the x-axis starts from 2 because one node in one community is not considered overlapping. The analysis also delves into an advanced feature, discussing the total number of groups considering the inclusion of overlapping nodes. It is observed that there is a negative correlation between the total number of groups and the presence of overlapping nodes. This implies that when nodes are assigned to multiple groups, the existence or validity of the community structure becomes questionable, particularly when the number of nodes in a community exceeds 6.

Overall, the analysis of the results highlights the characteristics of the smart healthcare network. It shows the distribution patterns of density, node numbers, and overlapped nodes, shedding light on the structure and trust relationships within the network. The presence of single-node communities suggests a lack of connections and trust, while the decreasing number of overlapped nodes indicates less overlap as nodes are assigned to more communities. These insights provide valuable information for understanding the community structure and dynamics within the smart healthcare network.

\subsection{Evaluations on the Security Level of Blockchain}

In this section, we discuss how the aforementioned attacks are resistant to a satisfyingly high degree by using the modified blockchain structure. To better compare the performance, we establish two personalized privacy protection models, one with the blockchain while the other without. The proposed blockchain structure is based on a Proof-of-Work (PoW) consensus algorithm. Thus, a large number of blocks means it is almost impossible to launch attacks because it cost too much hash rate. In the following context, we consider a start-up blockchain with a limited amount of blocks. In this case, the adversaries may have the incentive to mount relevant attacks.

\begin{figure}
\centering
\subfloat[Comparison of Models With and Without Blockchain]{
\includegraphics[width=2.25in]{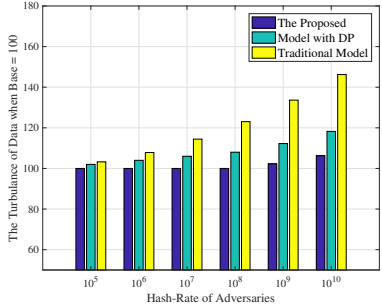}
\label{f_6a}
}
\hfil
\subfloat[Successful Attack Requirements with respect to the Number of Blocks.]{
\includegraphics[width=2.25in]{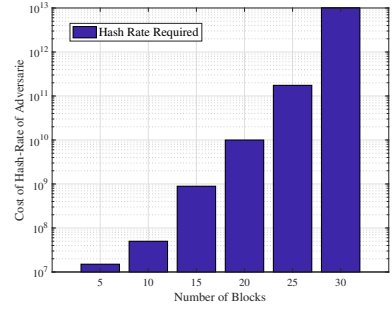}
\label{f_6b}
}
\caption{Performances comparison with Blockchain}
\label{f6}
\end{figure}

To compare the performances of the three models, we establish a coordinate system where the semi-logarithmic x-axis denotes the required hash rate and the linear y-axis is the turbulence of data when base=100. As shown in Fig.\ref{f_6a}, with the increase of the hash rate of the adversary, the turbulence grows correspondingly for all three cases. To successfully launch attacks, the hash rate of adversaries should pass a specific threshold. Usually, in PoW-based consensus blockchain systems, the threshold is believed to be $50\%$. That means if the hash rate of the adversary is over half of all hash rates, the attack success and the adversary takes control of the blockchain system. As indicated in our simulation, there are only $10$ blocks in total. Apparently, in practice, there are hundreds of, or even thousands of blocks, appending one after another, and it needs unimaginable amount of computing power to make the attacks happen.

In order to show how much hash rate is required for launching the attack, we establish a coordinate system where the semi-logarithmic y-axis denotes the required hash rate and the linear x-axis is the number of blocks. From Fig.~\ref{f_6b}, it is intuitive that the required hash rate grows in an exponential manner with the increase of block number (shows linearly because of the semi-logarithmic y-axis). It is worth mentioning that the hash rate demand breaks through $10^{13}$ when the block number is $30$. Usually, the high hash rate will demotivate most adversaries. Although there are adversaries with extremely high hash rates, they may not benefit from attacking such a blockchain. Moreover, in PoW-based consensus blockchains, the mining difficulty lifts with rounds, and thereby the protection is enhanced with more blocks appending to the blockchain.

The performance of three models is compared using a coordinate system where the x-axis represents the required hash rate (semi-logarithmic scale) and the y-axis represents the turbulence of data when the base is set to 100. The results depicted in Fig.\ref{f_6a} show that as the hash rate of the adversary increases, the turbulence of the data grows accordingly for all three cases. This indicates that to successfully launch attacks, the hash rate of the adversary must surpass a specific threshold. In most Proof-of-Work (PoW) based consensus blockchain systems, this threshold is commonly believed to be 50\%. This means that if the adversary's hash rate exceeds half of the total hash rates in the network, they can successfully attack and take control of the blockchain system. However, in the simulation presented, there are only 10 blocks in total, which is significantly smaller than the typical number of blocks in practical scenarios. It is important to note that in real-world scenarios with hundreds or even thousands of blocks appending one after another, launching successful attacks would require an unimaginable amount of computing power.

A coordinate system is established to demonstrate the required hash rate for launching an attack. The y-axis represents the required hash rate (semi-logarithmic scale) and the x-axis represents the number of blocks. From Fig.\ref{f_6b}, it is evident that the required hash rate increases exponentially as the number of blocks increases (appearing linearly due to the semi-logarithmic y-axis). It is worth mentioning that the hash rate demand surpasses $10^{13}$ when the number of blocks reaches 30. Typically, a high hash rate requirement discourages most adversaries. Even if there are adversaries with extremely high hash rates, they may not find it beneficial to attack such a blockchain. Additionally, in PoW-based consensus blockchains, the mining difficulty increases with each round, resulting in enhanced protection as more blocks are appended to the blockchain.

Overall, the analysis of the results highlights the relationship between the required hash rate, number of blocks, and the security of the blockchain system. Increasing the hash rate of the adversary leads to higher turbulence in the data and increases the risk of successful attacks. However, the practical feasibility of launching such attacks is severely constrained by the vast computing power required, especially in scenarios with a large number of blocks. The increasing hash rate demands with the number of blocks also act as a deterrent for adversaries, while the mining difficulty mechanism in PoW-based blockchains adds an additional layer of protection as more blocks are added.

\begin{figure}[h]
\centering
\includegraphics[width=3.25in]{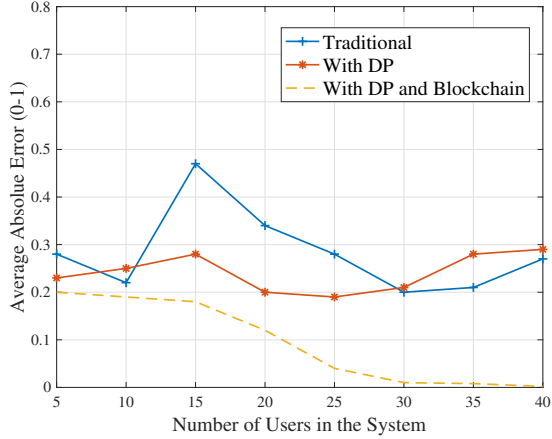}
\caption{Performance comparison with DP and Blockchain. \label{f10}}
\end{figure}

\subsection{Evaluations on the Blockchain Against Poisoning Attacks}

To validate the performances of poisoning attack proof, we compare the three models, in particular, classic differential privacy, personalized differential privacy, and blockchain-assisted personalized differential privacy. With 50 times experiments, the comparison of actual results and poisoned results are evaluated by average absolute error (AAE). In each round, we practically assume an attacker will poison 20\% of the whole dataset. Besides, the user number is used as the x-axis. From Fig.~\ref{f10}, if there are over $30$ users in a community, the AAE value of the blockchain-assisted model converges to $0$. This indicates the poisoning attack is totally mitigated. 

This is because of a practical assumption of blockchain, in particular, a \textit{sufficiently large community}. If an adversary hopes to poison the data on-chain, he/she needs over 50\% of the nodes to agree on through consensus. In most public blockchains running nowadays, there usually have a sufficiently large community to maintain the chain, for example, Ethereum. Usually, the assumption holds since the majority of the members are agreed on the same benefits. In the experiments, the nodes may become malicious with a chance of 50\% or 0\% before or after receiving any rewards. Then, we have the observation that the poisoning attacks are mitigated after the user number in a community passes $30$.

\subsection{Privacy Protection Measurement}

The privacy protection level is the most important index for privacy protection models. The most primary characteristic of personalized privacy protection is flexible and directional.

\begin{figure*}[!htbp]
\centering
\subfloat[]{
\includegraphics[width=2.25in]{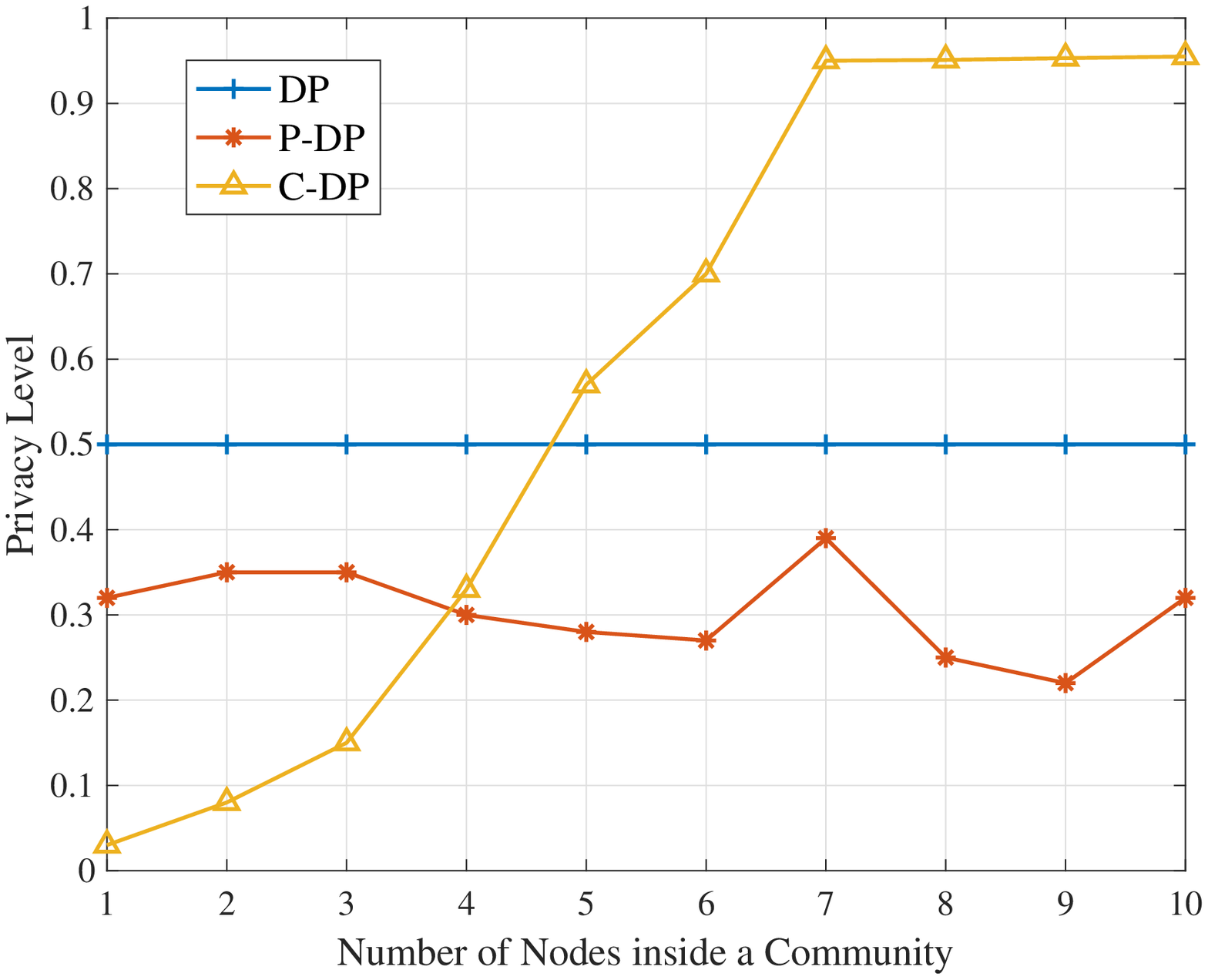}
\label{f_1aa}
}
\hfill
\subfloat[]{
\includegraphics[width=2.25in]{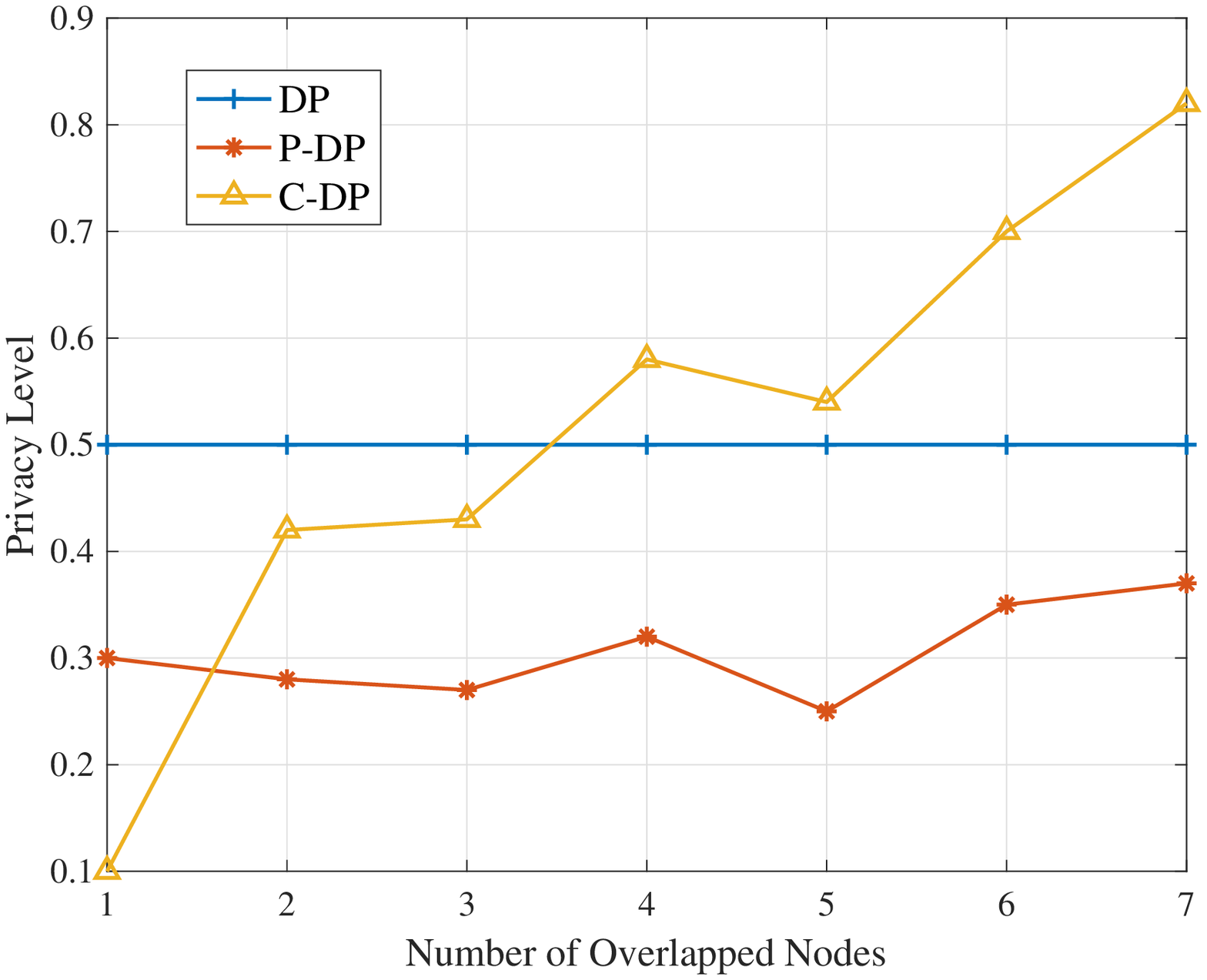}
\label{f_1ba}
}
\caption{Privacy Protection Measurement from Perspectives of Personalized Privacy and Overlapped Nodes}
\label{figure2}
\end{figure*}

In Fig.~\ref{f_1aa}, we can conclude that for DP, the privacy level maintains the same all the time. For P-DP, although the privacy level fluctuates a little bit, it is not practical because it takes all the users into consideration, including people who are not direct trustees. In C-DP, we only consider the users who are direct trustees. The direct friends are divided by the community density, and the privacy level complies with sigmoid function trends.

In Fig.~\ref{f_1ba}, the overlapped nodes play an important role as we have to decide the privacy level of overlapped nodes when they are in different groups with various densities. We provide the lowest level based on the communities it involves. Therefore, we can observe a fluctuation in the privacy protection level. If the node is inside five communities, but all the communities have a relatively low density, it may have a relatively high privacy level and vice versa.

The results depicted in Fig.~\ref{f_1aa} demonstrate the behavior of different privacy preservation approaches: DP, P-DP, and C-DP. In the case of DP (Differential Privacy), the privacy level remains the same throughout the entire duration. This indicates a consistent and fixed level of privacy protection, which may not be tailored to individual users' specific requirements. For P-DP (Personalized Differential Privacy), although the privacy level fluctuates slightly, it takes into account all users, including those who are not direct trustees. This approach may not be practical as it lacks precision in privacy level assignment. On the other hand, C-DP (Community-based Differential Privacy) focuses only on users who are direct trustees. The privacy level is determined based on the community density, and it follows the trends of a sigmoid function. This approach provides a more tailored and fine-grained privacy protection level, aligning with the characteristics of the communities and ensuring more personalized privacy preservation.

The analysis is centered around the role of overlapped nodes and their impact on privacy levels when they belong to different groups with varying densities. Fig.~\ref{f_1ba} presents the results. It is observed that the privacy protection level exhibits fluctuations due to the presence of overlapped nodes. The privacy level determination takes into account the communities in which the node is involved. If a node is part of multiple communities but all those communities have relatively low densities, it may have a higher privacy level. Conversely, if the node is involved in several communities with higher densities, it may have a lower privacy level. This fluctuation reflects the consideration of community characteristics and the varying influence of overlapping memberships on privacy preservation.

Overall, the analysis of the results highlights the differences and implications of various privacy preservation approaches. DP provides a fixed level of privacy, P-DP lacks precision, and C-DP offers personalized privacy levels based on community densities. Additionally, the presence of overlapped nodes introduces variations in privacy levels depending on the densities of the communities they belong to. These insights emphasize the importance of tailoring privacy preservation to individual users' needs and considering the dynamics of community structures in determining privacy levels for optimal privacy protection.

\subsection{Data Utility Comparison}

Data utility directly affects the quality of service of users. Therefore, we can not sacrifice too much data utility to achieve additional privacy load. Personalized privacy protection provides flexible data utility to different users, which can provide high-quality service to specific users.

Fig.\ref{f_2a} leverages stairs the chart to show the data utility increases with the increase of the $\epsilon$,  which is also known as privacy level. In terms of the overlapped nodes data utility in Fig.~\ref{f_2b}, we can see the fluctuating trend is similar to the privacy protection level as well. Thus, the trade-off between privacy protection level and data utility is nicely derived.

Fig.\ref{f_2a} presents a stair chart illustrating the relationship between data utility and the privacy level (represented by $\epsilon$). The chart demonstrates that as the privacy level (i.e., $\epsilon$) increases, the data utility also increases. This indicates that a higher level of privacy protection (achieved by increasing $\epsilon$) is associated with a higher level of data utility. The stair-like pattern suggests that there are distinct increments in data utility as the privacy level is adjusted, rather than a continuous and smooth progression. This implies that privacy preservation mechanisms are capable of striking a balance between protecting privacy and maintaining useful data.

Fig.\ref{f_2b} depicts the data utility of overlapped nodes. The fluctuating trend observed in the data utility aligns with the privacy protection level. This implies that as the privacy level varies (e.g., influenced by community densities or specific privacy mechanisms), the data utility of overlapped nodes follows a similar fluctuating pattern. The trade-off between privacy protection level and data utility is clearly demonstrated, suggesting that increasing privacy protection measures may come at the cost of some loss in data utility. However, the fluctuating trend indicates that there might be opportunities to optimize the privacy-utility trade-off by fine-tuning the privacy mechanisms or adjusting community characteristics.

Overall, the analysis of the results highlights the relationship between privacy protection level, data utility, and the influence of overlapped nodes. It reveals that increasing the privacy level tends to improve data utility, indicating that privacy preservation measures can be designed to strike a balance between privacy protection and maintaining valuable data. The fluctuating trend in data utility and its similarity to the privacy protection level emphasizes the trade-off between privacy and utility, presenting opportunities for further optimization in privacy mechanisms. These findings contribute to understanding the interplay between privacy and data utility in the context of the depicted scenarios.

\begin{figure*}[!htbp]
\centering
\subfloat[]{
\includegraphics[width=2.25in]{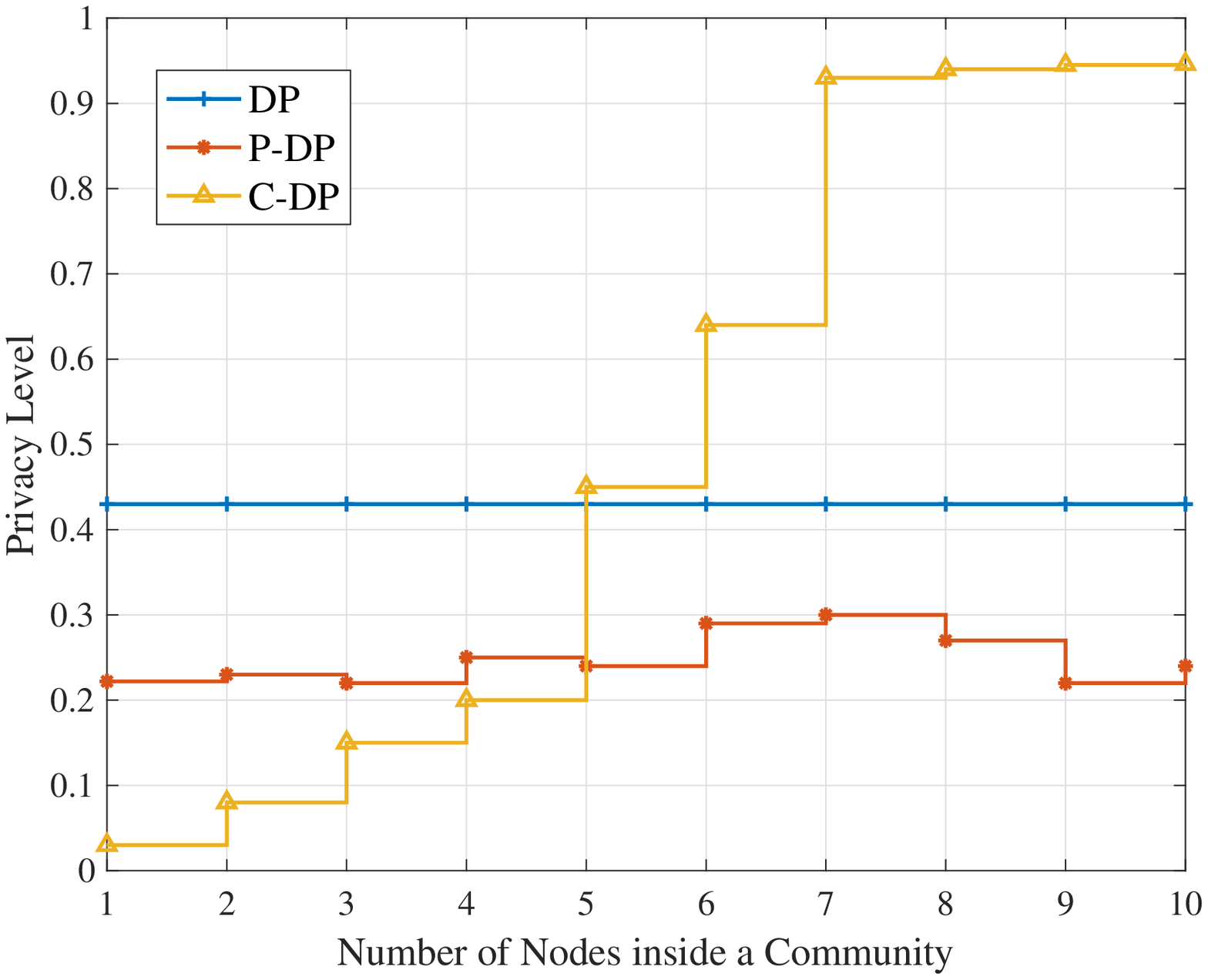}
\label{f_2a}
}
\hfill
\subfloat[]{
\includegraphics[width=2.25in]{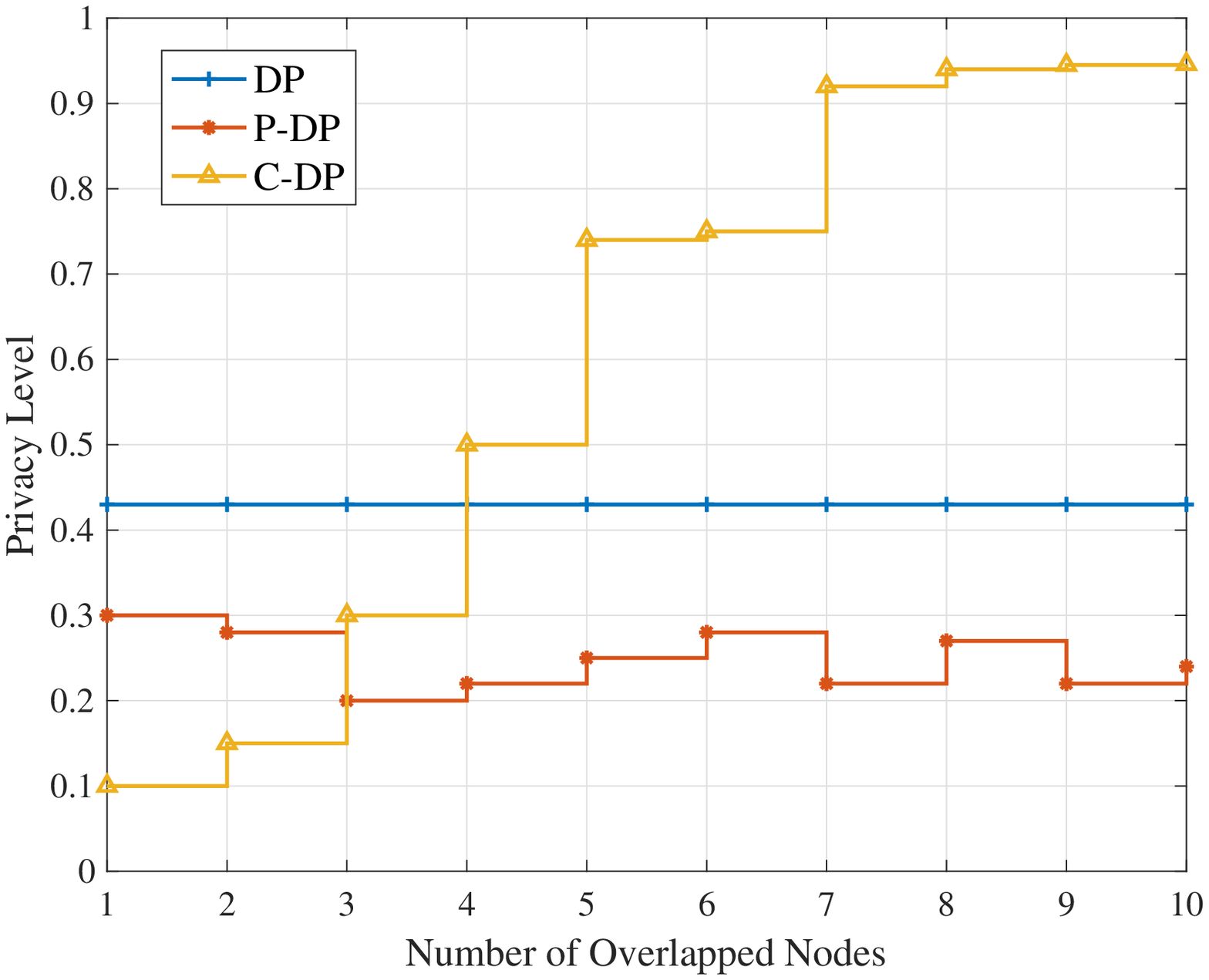}
\label{f_2b}
}
\caption{Data Utility Measurement from Perspectives of Personalized Privacy and Overlapped Nodes}
\label{figure3}
\end{figure*}

\subsection{Evaluation on the Computation Overhead of Optimization for Tradeoff}

\begin{figure}[h]
\centering
\includegraphics[width=3.25in]{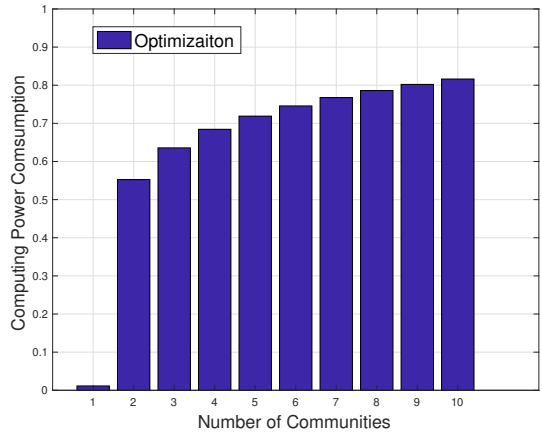}
\caption{Evaluation on the Computation Overhead of Optimization for Tradeoff}
\label{f_opt}
\end{figure}

In Fig.~\ref{f_opt}, we have demonstrated the computation overhead of the optimization for the tradeoff. We evaluated the change of computation overhead regarding the number of communities inside a healthcare network. Since there is a personalized mapping function, the increment of a number of communities costs less and less computation power. If the number of communities is great enough, the computation overhead will converge to a specific value, which testifies the scalability of the proposed model in this big data era.

\section{Concluding Remarks}
\label{sec:summary}

In this article, we start by describing the vulnerabilities of SHNs, including privacy leakages concerns, trade-off issues, as well as linkage and poisoning attacks. To solve the problems, we devise a personalized and trustworthy privacy protection model considering the trust measured by community density. By clearly defining the community, we then personalize the privacy protection levels constrained by differential privacy. To mitigate linkage attacks, a noise correlation uncoupling mechanism is proposed. In this case, even conducting linkage attacks, no more sensitive information can be obtained, which wipes out the incentive of attackers. Meanwhile, the integration of blockchain can defeat data falsification attacks. We perform corresponding experiments and the results confirm its effectiveness.

It is worth noting that in the proposed structure, the community detection is an important procedure but not easy to control the control the granularity. In addition, the theoretical foundation of the optimized trade-off should be further clarified.

Future work in progress includes establishing the model using game theory, which can better describe the confrontation of data holders and adversaries. This will help with deriving a better balance between privacy and data utility. Besides, the integration of federated learning to enhance privacy protection in this scenario is ongoing.

\section*{Acknowledgement}

This paper is in part funded by the National Key Research and Development Program of China (2021YFF0900400).


\bibliographystyle{plain}
\bibliography{mybibfile}

\vspace{15mm}


{\includegraphics[width=1in,height=1.25in,clip,keepaspectratio]{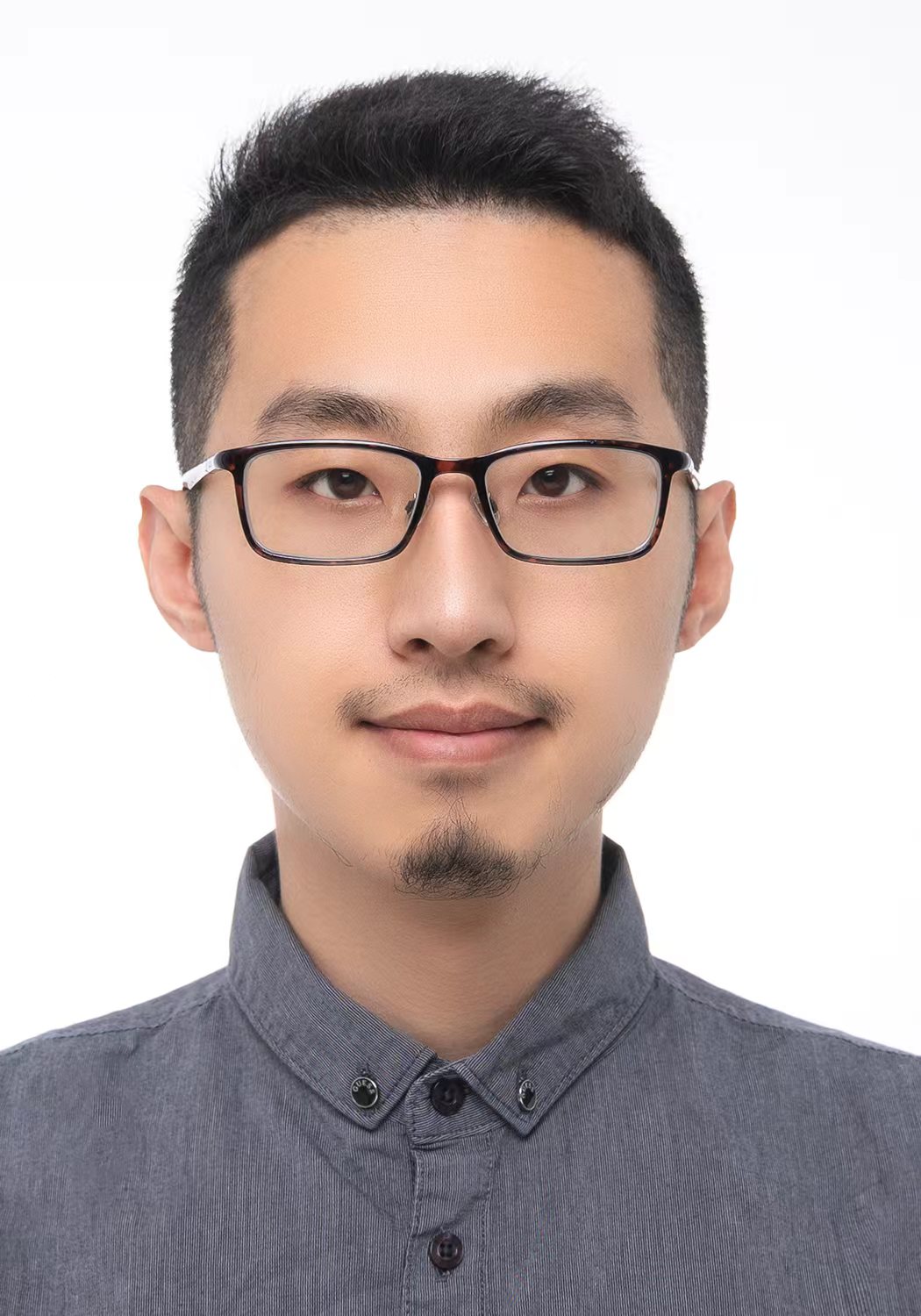}}{Youyang Qu} is currently a research scientist of data61, Commonwealth Scientific and Industrial Research Organization (CSIRO), Australia. Before joining CSIRO, he served as a research fellow at Deakin University. He received his B.S. degree in Mechanical Automation in 2012 and M.S. degree in Software Engineering in 2015 from the Beijing Institute of Technology, respectively. He received his Ph.D. degree at the School of Information Technology, Deakin University, in 2019. His research interests focus on Machine Learning, Big Data, IoT, blockchain and corresponding security and customizable privacy issues. He has over 50 publications, including high-quality journals and conference papers such as IEEE TII, IEEE TNSE, ACM Computing Surveys, IEEE IOTJ, etc. He is active in the research society and has served as an organizing committee member in SPDE 2020, BigSecuirty 2021, Tridentcom 2021/2022.

\vspace{5mm}

{\includegraphics[width=1in,height=1.25in,clip,keepaspectratio]{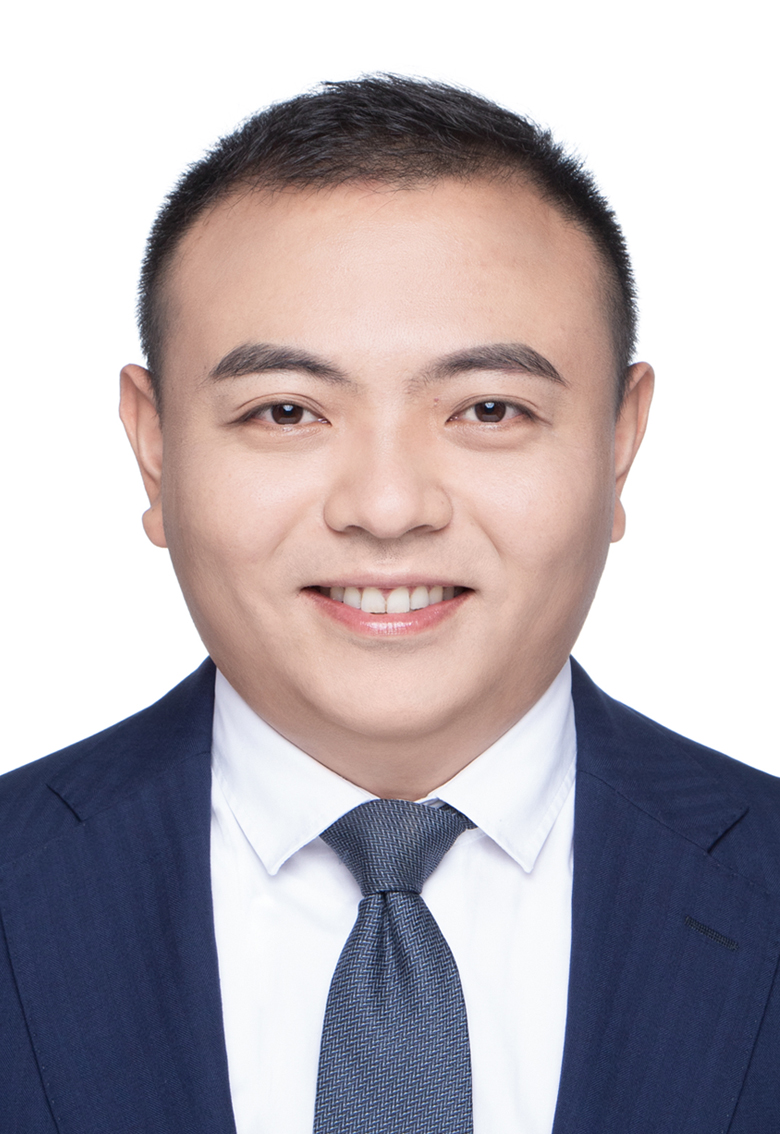}}{Lichuan Ma} received his B.S. degrees in Information Security from the School of Mathematics, Shandong University, in 2012 and the Ph.D degree in information security from Xidian University, in 2018. Now, he is with the School of Cyber Engineering at Xidian University and a member of Shaanxi Key Laboratory of Blockchain and Secure Computing. His research interests focus on trust management and privacy-preserving techniques for intelligent systems.

\vspace{5mm}

{\includegraphics[width=1in,height=1.25in,clip,keepaspectratio]{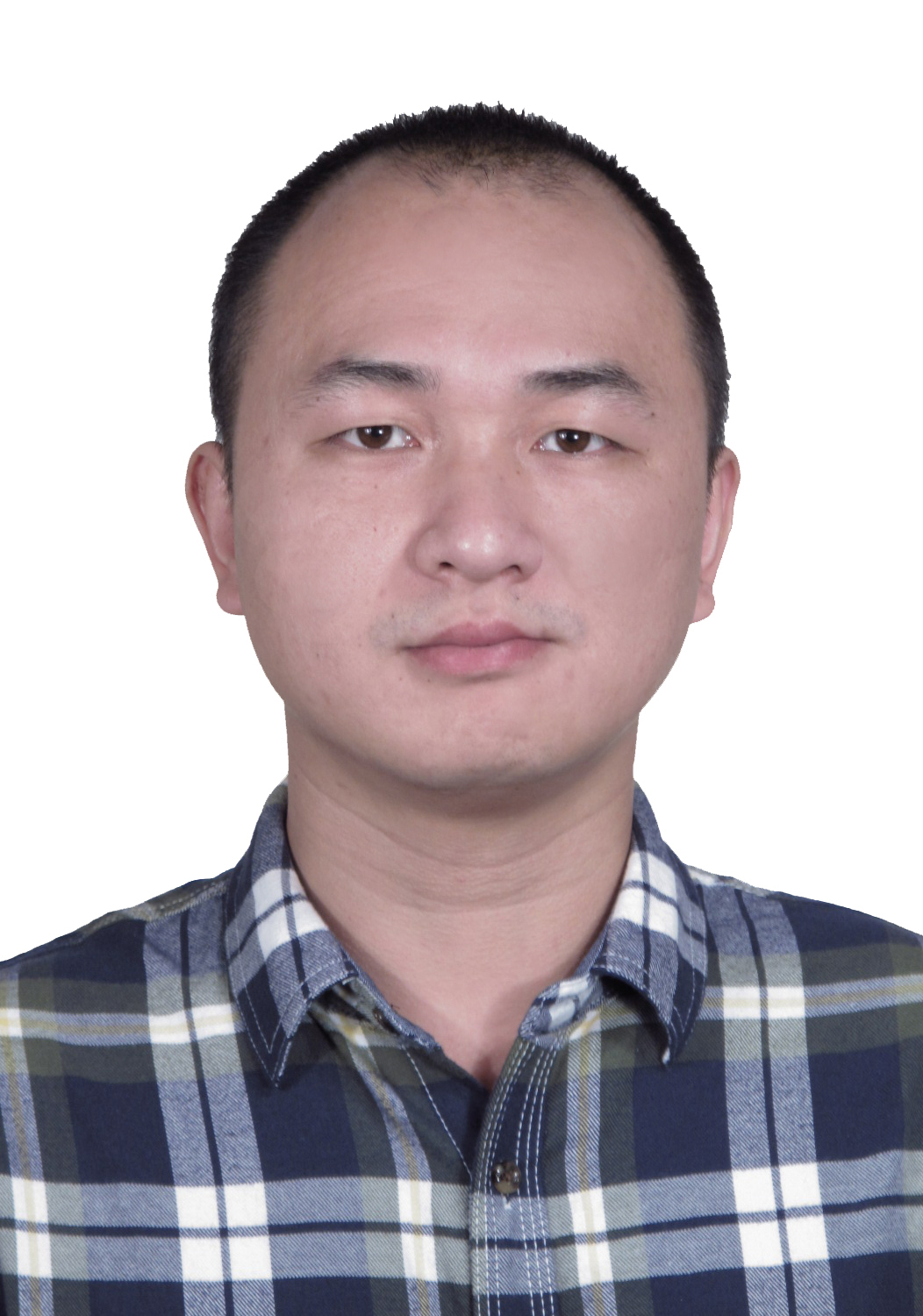}}{Wenjie Ye} received his B.S. degrees in Information Security from the School of Mathematics, Shandong University, in 2012 and the Ph.D degree in information security from Xidian University, in 2018. Now, he is with the School of Cyber Engineering at Xidian University and a member of Shaanxi Key Laboratory of Blockchain and Secure Computing. His research interests focus on trust management and privacy-preserving techniques for intelligent systems.

\vspace{5mm}

{\includegraphics[width=1in,height=1.25in,clip,keepaspectratio]{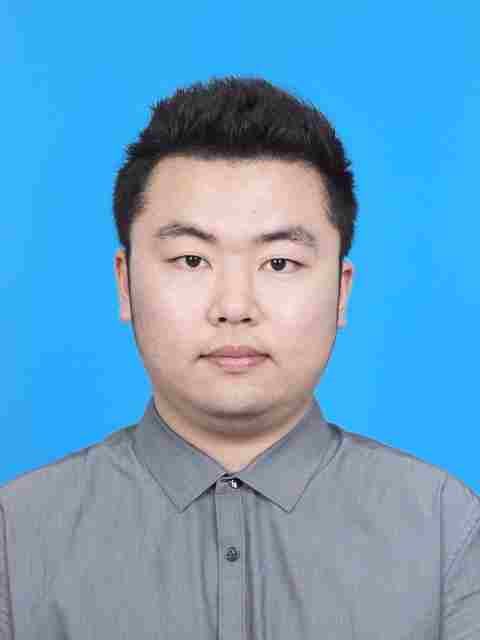}}{Xuemeng Zhai} received his B.S. degrees in Information Security from the School of Mathematics, Shandong University, in 2012 and the Ph.D degree in information security from Xidian University, in 2018. Now, he is with the School of Cyber Engineering at Xidian University and a member of Shaanxi Key Laboratory of Blockchain and Secure Computing. His research interests focus on trust management and privacy-preserving techniques for intelligent systems.

\vspace{5mm}

{\includegraphics[width=1in,height=1.25in,clip,keepaspectratio]{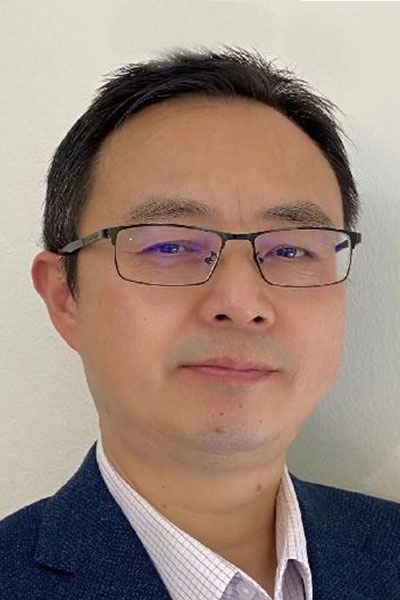}}{Shui Yu} obtained his PhD from Deakin University, Australia, in 2004. He currently is a Professor of School of Computer Science, University of Technology Sydney, Australia. Dr Yu’s research interest includes Network Science, Security and Privacy, Big Data, and Mathematical Modelling. He has published five monographs and edited two books, more than 500 technical papers at various venues, such as IEEE TPDS, TDSC, TIFS, TC, TMC, TKDE, TETC, ToN, and INFOCOM. His h-index is 66. Dr Yu promoted the research field of networking for big data in 2013, and his research outputs have been widely adopted by industrial systems, such as Amazon cloud security. He is passionate for professional services for his communities. He is currently serving a couple of prestigious editorial boards, including IEEE Communications Surveys and Tutorials (Area Editor), IEEE Internet of Things Journal, and so on. He served as a Distinguished Lecturer of IEEE Communications Society (2018-2021), and is a Distinguished Lecturer of IEEE Vehicular Technology Society, a Distinguished Visitor of IEEE Computer Society, a member of IEEE ComSoc Educational Services Board, an elected member of Board of Governors of IEEE ComSoc and VTS. He is a member of ACM and AAAS, a Fellow of IEEE.

\vspace{5mm}

{\includegraphics[width=1in,height=1.25in,clip,keepaspectratio]{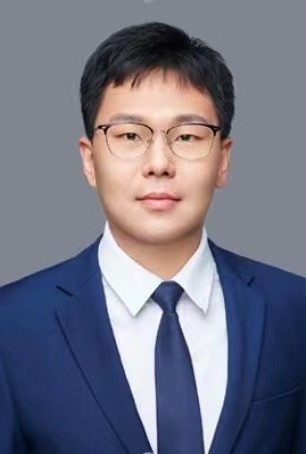}}{Yunfeng Li},  Ph.D. in computer simulation and chemical engineering from Monash University, Australia, won the honorary title of the young backbone talent in Beijing. He is currently the CEO of CNPIEC KEXIN LTD. He has won a scholarship from the University of New South Wales in Australia and a scholarship from Monash University in Australia. He is mainly engaged in the research of big data, artificial intelligence, computer simulation and other directions, and leads the industrial application based on artificial intelligence technology.

\vspace{5mm}

{\includegraphics[width=1in,height=1.25in,clip,keepaspectratio]{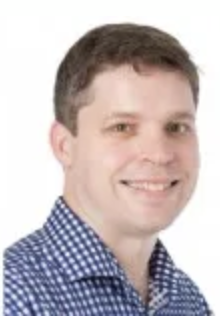}}{David Smith} is a Principal Research Scientist with Data61, CSIRO (previously in NICTA), leading the Distributed Privacy and Security team in the Information Security and Privacy group of the Software and Computational Systems Program. He was with NICTA since 2004, and in Data61 since 2016. David received the B.E. degree in electrical engineering from the University of New South Wales, Australia, in 1997, and the M.E. (research) and Ph.D. degrees in telecommunications engineering from the University of Technology Sydney, in 2001 and 2004 (graduating in 2005), respectively. David has published over 150 technical refereed papers and made various contributions to IEEE standardization activity, including being listed as a major contributor to IEEE 802.15.8. His research interests are in data privacy, distributed systems privacy and edge computing data privacy, distributed machine learning, data privacy for supply chains, wireless body area networks, game theory for distributed networks, 5G networks, disaster tolerant networks, and distributed optimization for smart grid.

\end{document}